# Pollen Patterns Form from Modulated Phases

Asja Radja[1], Eric M. Horsley[1], Maxim O. Lavrentovich[2*], and Alison M. Sweeney[1*]

asjar@sas.upenn.edu
ehorsley@sas.upenn.edu
*mlavrent@utk.edu
*alisonsw@sas.upenn.edu

[1]Department of Physics and Astronomy, University of Pennsylvania, 209 S. 33rd St., Philadelphia, Pennsylvania 19104, U.S.A.
[2]Department of Physics and Astronomy, University of Knoxville, 1408 Circle Dr., Knoxville, Tennessee 37996, U.S.A

*Co-corresponding authors

**Abstract**

Pollen grains are known for their impressive variety of species-specific, microscale surface patterning. Despite having similar biological developmental steps, pollen grain surface features are remarkably geometrically varied. Previous work suggests that a physical process may drive this pattern formation and that the observed diversity of patterns can be explained by viewing pollen pattern development as a phase transition to a spatially modulated phase. Several studies have shown that the polysaccharide material of plant cell walls undergoes phase separation in the absence of cross-linking stabilizers of the mixed phase. Here we show experimental evidence that phase separation of the extracellular polysaccharide material (primexine) during pollen cell development leads to a spatially modulated phase. The spatial pattern of this phase-separated primexine is also mechanically coupled to the undulation of the pollen cell membrane. The resulting patterned pools of denser primexine form the negative template of the ultimate sites of sporopollenin deposition, leading to the final micropattern observed in the mature pollen. We then present a general physical model of pattern formation via modulated phases. Using analytical and numerical techniques, we find that most of the pollen micropatterns observed in biological evolution could result from a physical process of modulated phases. However, an analysis of the relative rates of transitions from states that are equilibrated to or from states that are not equilibrated suggests that while equilibrium states of this process have occurred throughout evolutionary history, there has been no particular evolutionary selection for symmetric, equilibrated states.

**Introduction**

The diversity and beauty of pollen grain surface patterns have intrigued scientists for decades, yet no unifying theory has emerged to explain either the pattern formation mechanism or the function of these surface features (Fig. 1)[1]. So, a natural question is: how do pollen grains create such diverse, microscale patterns when other cells typically do not? Has there been evolutionary selection for symmetric patterns, or are these patterns the result of evolutionary drift of a separate biochemical process? Geometrically similar patterns are found on fungal spores, mite carapaces, and insect eggs, but these patterns are not nearly as diverse as those found on pollen[2]. The multitude of pollen patterns observed in nature, along with a complex extracellular composition, make understanding pollen development particularly difficult. Our objective is to provide a unified conceptual framework for understanding the patterning process.

In mature pollen, the outermost layer of the extracellular material is highly patterned and called the exine. The exine is a chemically and physically robust outer wall made of sporopollenin, a complex, highly resistant chemical whose structure and composition are not fully described[3]. Apart from the structure of the exine itself, pollen can be patterned with a varying number and geometric arrangement of apertures, which are regions of the extracellular material that have a reduced or absent exine and are the sites where the pollen tube emerges during germination[4]. Apertures also allow the pollen grain to reversibly fold during desiccation and rehydration[5].

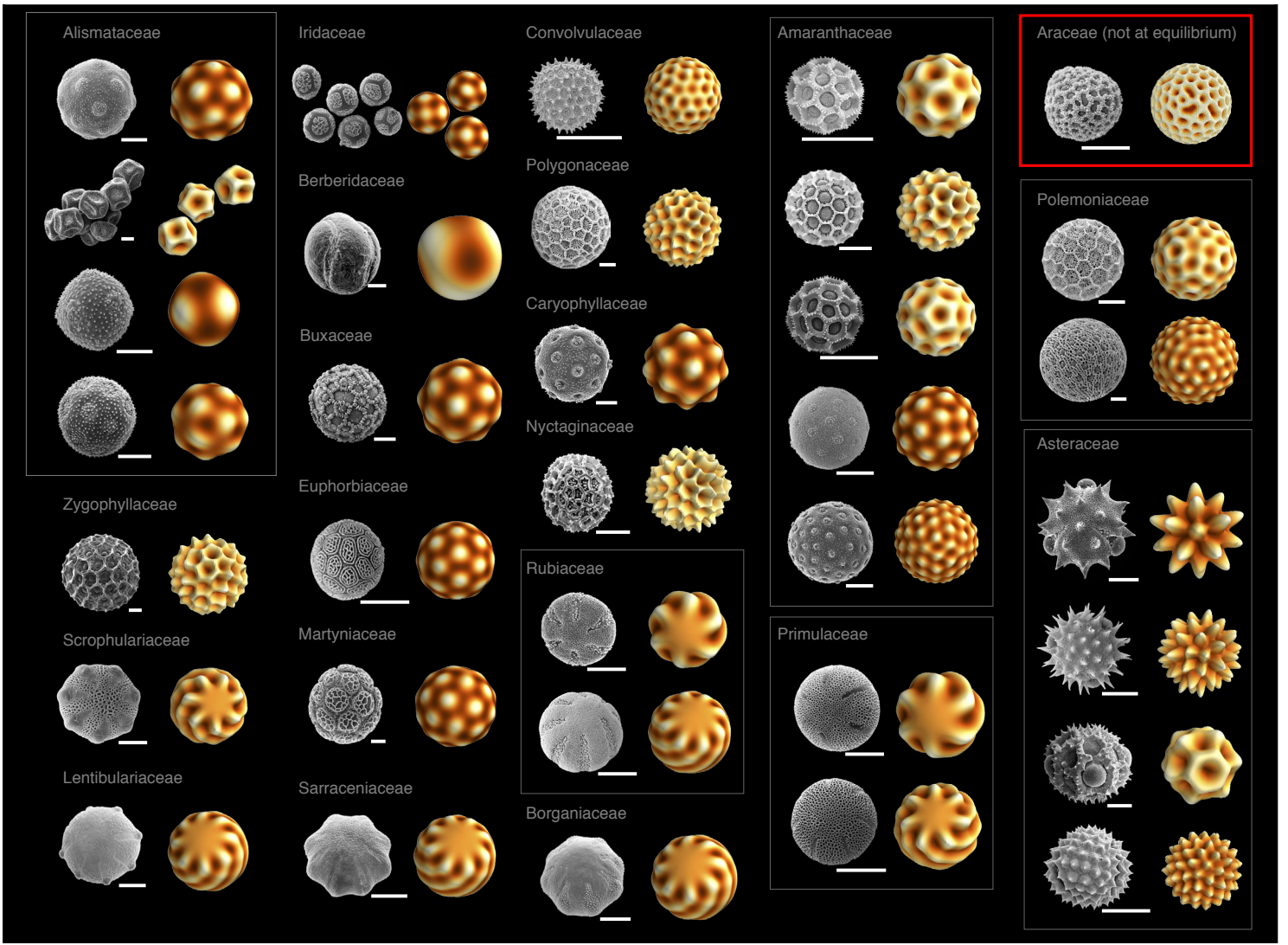

**Figure 1: Pollen SEMs and recapitulated patterns.** Pairs of images illustrate examples of pollen grain surface patterns reproduced with our simulations. These selected pairs represent examples of the range of patterns we found at equilibrium patterns (polygonal spikes, polygonal holes, chiral stripes, and mixtures of these); the red box represents an example of non-equilibrium patterns that are mostly foamy (reticulate). The left image in each pair shows the SEM of a given species, the right image in each pair shows the simulated surface of the same geometry. The species represented and the Hamiltonian parameters ($l_0, \lambda'_3$) producing the matching simulated surface are listed. All SEM micrographs from paldat.org; all equilibrium simulations used $(R^2\tau)/K = -1$. **First column:** *Alisma lanceolatum*, (9.5,1); *Caldesia parnassifolia*, (5.5,-1); *Echinodorus cordifolius*, (4.5,1); *Echinodorus quadricostatus*, (7.5,1); *Kallstroemia maxima*, (16.5,-1); *Diascia barberae*, (12.5,0); *Utricularia sandersonii*, (13.5,0). **Second column:** *Iris bucharica*, (6.5,1); *Berberis vulgaris*, (3.5,0); *Sarcococca hookeriana*, (9.5,1); *Phyllanthus sp.*, (11.5,1); *Ibicella lutea*, (11.5,1); *Sarracenia flava*, (13.5,0). **Third column:** *Ipomoea cholulensis*, (19.5,-1); *Persicaria mitis*, (19.5,-1); *Cerastium tomentosum*, (9.5,1); *Bougainvillea sp.* (17.5,-1); *Galium wirtgenii*, (8.5,0); *Galium album*, (13.5,0); *Arnebia pulchra*, (13.5,0). **Fourth column:** *Pfaffia gnaphaloides*, (10,-1); *Gomphrena globosa* (15.5,-1); *Pfaffia tuberosa*, (12,-1); *Amaranthus blitum*, (12.5,1); *Chenopodium album*, (19.5,1); *Primula veris*, (8.5,0); *Primula elatior*, (12.5,0). **Fifth column:** (red box) *Anthurium gracile*, simulated with conserved dynamics [Eq. (7) with $D=K=1, q_0=1.5, \tau=-20, \lambda_3=-20, \lambda_4=120$, and a sphere radius of $R=15$, for dimensionless parameters $l_0=22.5, \lambda_3' \approx -27.4, (R^2\tau)/K=-4500$. We used a Gaussian, random initial $\psi(\theta,\varphi)$ centered around 0 with a variance of 0.04, and evolved the field until time t=2]; *Phlox drummondii*, (16.5,-1); *Polemonium pauciflorum*, (20.5,1); *Gaillardia aristata*, (10,1); *Bidens pilosa*, (14.5,1); *Chondrilla juncea*, (8,-1); and *Iva xanthiifolia*, (19.5,1).

The general developmental steps that result in the observed variety of pollen surface patterns are well-characterized[6]. The cell wall of the meiotic mother cell fails to completely divide, leaving the resulting daughter pollen cells contained within a specialized structure called the callose wall, and, as a result, they are isolated from the rest of the anther fluid. The callose wall has an unusual composition of β-1,3 glucan, which provides an experimental strategy for its selective degradation to access the developing pollen grains[7]. The developing pollen cells then secrete a polysaccharide material called the "primexine" to the cell surface; the primexine accumulates between the cell's plasma membrane and the callose wall. The composition of the primexine is not well-characterized but is likely to be a high molecular weight polysaccharide[8]. It has been established that the global pattern features of the mature pollen wall (exine) are somehow templated by the developing primexine layer during this enclosed "tetrad" stage[8,9,10], though the physical mechanism of this process remains undescribed. Following this global templating by the primexine, the callose wall dissolves and sporopollenin is secreted by adjacent tapetal cells and accumulates on the pollen cell surface, resulting in the patterned exine layer of mature pollen (Fig. 1).

Several studies suggest that pollen apertures may also be features dictated by the primexine process, especially in multi-aperturate and spiraperturate pollen[11,12]. However, in pollen grains that contain fewer than six apertures, the aperture pattern may be established by points of cellular contact between daughter cells during meiosis[13]; since apertures possibly arising by this mechanism have a tetrad geometry of daughter cells, they are easy to identify, and we excluded them from this analysis of pattern formation via a primexine template. It is clear that there is no one unified cell developmental mechanism of aperture formation across plants[11]; therefore, we adopt the definition that apertures are simply thin regions of the exine material. Here, we provide a physical explanation for the generation of the templated pattern by the primexine material.

A physical theory for cell surface patterning via a first-order phase transition of material deposited on the cell surface was recently reported by Lavrentovich and colleagues[14]. Here we treat the primexine as a phase-separating concentration field on a spherical surface, which in turn introduces heterogeneities (e.g., a locally varying pressure or preferred curvature) and a local buckling of the plasma membrane. Such heterogeneities, when coupled to the elasticity of a membrane, are known to create spatially modulated structures[15]. In pollen, a mechanical coupling between the polysaccharide matrix and membrane may be promoted by the presence of the outer callose wall that encapsulates extracellular polysaccharides near the cell membrane during pattern formation. Initial pattern formation could then occur via a phase transition of the polysaccharide to a spatially modulated state. The same kind of transition has been used to describe the formation of viruses[16,17] and two-component vesicles[18,19], which are also intricately patterned spherical objects, and discretized versions of such patterned spherical objects have also been computationally explored[20]. We employ a fully spectral method that allows for a systematic characterization of pattern configurations.

In addition to the pollen pattern formation process being unknown, there has been no unifying, satisfactory answer to what the functional role of these patterns might be, in spite of many previous efforts. Some studies have found a correlation between pollinator types and pollen grain surface features[21]. Other studies have found that there is a general trend of increasing aperture number in angiosperms[22]. However, the findings of these studies often conflict, and there is no current consensus as to which features of pollen patterns may be evolutionarily selected for and why.

We show that the preponderance of extant pollen patterns can be explained through a phase transition of the primexine coupled to the plasma membrane during cell development. We also show novel experimental corroboration of a densification and pooling of primexine material leading to membrane undulations at the wavelength of the mature pollen pattern in *Passiflora incarnata*, a species whose exine is reticulate (foamy). This mechanism implies that evolutionary pattern diversity is to be expected, given the general chemical composition and physical makeup of the pollen grain during development and that the spherical surface of pollen grains must accommodate spherical defects in the resulting pattern. Further, most of the ordered states observed in evolved pollen pattern diversity can be recapitulated with a unique set of parameters in our theory (Fig. 1). Our theory is also able to account for patterns generated by this physical mechanism that do not reach an energy minimum (Fig. 1, red box). A surprise in our results is that the majority of mature, extant pollen patterns do not exist at energy minima within this pattern formation landscape; there apparently has been no strong evolutionary selection for symmetry via pattern equilibration in pollen. Finally, we propose a new way of characterizing pollen patterns motivated by this physical theory that is grounded in the physiology of pollen development.

**Materials and Methods**
*Microscopy*

*Passiflora incarnata* (Shady Oak Butterfly Farm) was grown at the University of Pennsylvania Department of Biology greenhouse under a 16 hour/day light cycle at a mean temperature of 77°F. Fresh anthers were collected, and pollen was immediately dissected out of the developing anthers within flower buds. To identify the stage of pollen development in a given anther, one anther from each flower bud was pressed between glass slides and examined with a brightfield optical microscope; only pollen in the tetrad stage was kept for further analysis.

For transmission electron microscopy (TEM), anthers were first fixed in 3% gluteraldehyde with 1% alcian blue in 1x phosphate-buffered saline (PBS) for 24 hours[23], and then post-fixed in 2% osmium tetroxide for 30 minutes. Next, an ethanol dehydration series was performed, and samples were embedded in Spurr's resin. Transverse ultrathin sections of 70 nm were cut with a Diatome diamond knife on a Reichert Ultracut-S microtome. Secondary staining was done with uranyl acetate and lead citrate. Sections were placed on copper mesh grids and imaged with a JEOL JEM-1010 electron microscope.

For scanning electron microscopy (SEM), we first separated the developing tetrads from their anthers and then enzymatically removed the callose walls, as described by Kirkpatrick and Owen[24]. The pollen grains from a single developing flower were placed in 1mL of 0.3% w/v cellulase, pectolyase and cytohelicase, 1.5% sucrose, and 1% polyvinylpyrolidone for 2 hours (Sigma-Aldrich; Milwaukee, MI). Next, the pollen grains were fixed in 3% gluteraldehyde in 1x PBS for 1 hour. Samples were then washed in deionized water for 5 minutes and placed in handmade Nitex bags (1 cm²); the bags were then heat sealed. The bags with the pollen samples were then submerged in 1x PBS for 5 minutes, followed by an ethanol dehydration series. Samples were then critical-point dried in $CO_2$ in a Tousimi Autosamdri-850. The pollen grains were removed from the bags, placed onto SEM stubs and sputter coated with a ~10 nm thick layer of gold-palladium using an SPI Module Sputter Coater. We prepared pollen grains at the same stage without enzymatically removing the callose walls as control for any unintended effects of the removal procedure. Samples were imaged using a FEI Quanta FEG 250.

*Primexine composition*

Pollen grains at the tetrad stage were also collected to analyze their primexine composition. We dissected pollen grains from anthers and enzymatically removed the callose walls using the method described in the section above. The whole pollen grains (without their callose walls) were then frozen and shipped over dry ice to the Complex Carbohydrate Research Center at the University of Georgia for a glycosyl composition and linkages analysis. The monosaccharide composition and linkages analyses were performed by combined gas chromatography/mass spectrometry of the per-*O*-trimethylsilyl derivatives as described previously by Santander and colleagues[25]. More details on the method used are in the supplemental information.

*Theoretical Model*

We describe the formation of the pollen surface pattern as a phase separation of the primexine mechanically coupled to the underlying plasma membrane. It should be noted that we are not modeling any detailed material properties of the primexine, but we do assume that it is able to phase separate, similar to mixtures of other high molecular weight extracellular polysaccharides such as hemicellulose and pectin[26,27]. This model is described in more detail in a previous study where the effects of thermal fluctuations on patterned states were additionally considered[14]. The present work focuses on a microscopic model without these fluctuation effects to study the number, variety, and stability of ordered states (which is much more difficult to do in the fluctuating case). We will include a brief description here for clarity.

Consider a scalar field, $\psi$, which represents the concentration field of the primexine polysaccharides in contact with the outer surface of a pollen grain plasma membrane. We postulate that the phase separation of this material drives the pattern formation of the pollen surface. The general Landau-Ginzburg free energy for $\psi$ is given by

$$\mathcal{H}[\psi] = \int d^2x \left[ \frac{\kappa_0}{2} |\nabla \psi|^2 + \frac{\tau_0}{2} \psi^2 + \frac{\lambda_3}{3!} \psi^3 + \frac{\lambda_4}{4!} \psi^4 \right], \tag{1}$$

where $\kappa_0$ and $\lambda_{3,4}$ are constants that depend on some undefined primexine chemical or material properties. We assume that $\kappa_0, \lambda_4 > 0$, and $\tau_0$ is a temperature-like term that is quenched below some critical value during pattern formation. Because this field sits on a spherical surface, we use spherical coordinates, $\psi = \psi(\theta, \phi)$ and our integration measure reads $\int d^2x = R^2 \int d\theta d\phi$, where $\theta \in [0, \pi]$ and $\phi \in [0, 2\pi)$. We then expand $\psi(\theta, \phi)$ in terms of spherical harmonics, $Y_l^m(\theta, \phi)$:

$$\psi(\theta,\phi) = \sum_{l=0}^{\infty}\sum_{m=-l}^{l} c_l^m Y_{lm}(\theta,\phi) \equiv \sum_{l} c_l^m Y_{lm} \tag{2}$$

Finally, the expansion coefficients satisfy the property $[c_l^m]^* = (-1)^m c_l^{-m}$ because the scalar field $\psi$ is real.

We now follow the infinite flat membrane analogue of our model studied by Leibler and Andelman[15]. Non-patterned (uniform) states, $l=0$ modes, are preferred in Eq (1). However, when we couple this field, $\psi$, to the local membrane curvature, we observe patterned states. The $l \neq 0$ modes are more energetically favorable in the coupled system since the primexine concentration on the surface causes the membrane to bend and fluctuate away from a spherical shape. Details of the implementation of this coupling are described in the work done by Lavrentovich and colleagues[14]. The salient physical feature of coupling the primexine to the cell membrane in this study is that a spatially modulated phase with a characteristic mode number $l_0$ arises, which approximately describes the number of times a given pattern wraps around the sphere. It is related to the characteristic wavelength, $\lambda$, by $l_0 \approx 2\pi R/\lambda$. The effective free energy for the field near $l \approx l_0$ has the general form

$$\mathcal{H} = \frac{1}{2}\sum_{l}[K(l-l_0)^2 + R^2\tau]|c_l^m|^2 + \mathcal{H}_{\text{int.}} \tag{3}$$

where $K$ and $\tau$ are new constants that depend on the material properties of the primexine and various physical parameters of the plasma membrane such as bending rigidity, surface tension, elasticity and/or lipid/protein density. These parameters may also incorporate features of the callose wall if the wall participates in inducing the membrane buckling. The terms in $\mathcal{H}_{\text{int.}}$ are inherited from Eq. (1) and involve couplings between different spherical harmonics:

$$\mathcal{H}_{\text{int.}} = \frac{R^2 \lambda_3}{3!}\Upsilon_{m_1,m_2,m_3}^{l_1,l_2,l_3} c_{l_1}^{m_1} c_{l_2}^{m_2} c_{l_3}^{m_3} + \frac{R^2 \lambda_4}{4!}\Upsilon_{m_1,m_2,\overline{m}}^{l_1,l_2,\bar{l}}\Upsilon_{m_3,m_4,-\overline{m}}^{l_3,l_4,\bar{l}} c_{l_1}^{m_1} c_{l_2}^{m_2} c_{l_3}^{m_3} c_{l_4}^{m_4} \tag{4}$$

where the $\Upsilon$s are Gaunt coefficients, with sums implied on all indices. Written in terms of the Wigner-3j symbols[28], the Gaunt coefficients are given by

$$\Upsilon_{m_1,m_2,m_3}^{l_1,l_2,l_3} \equiv \sqrt{\frac{\prod_{i=1}^{3}(2l_i+1)}{4\pi}}\begin{pmatrix} l_1 & l_2 & l_3 \\ 0 & 0 & 0 \end{pmatrix}\begin{pmatrix} l_1 & l_2 & l_3 \\ m_1 & m_2 & m_3 \end{pmatrix}. \tag{5}$$

Rapid evaluation algorithms are available for these symbols[29] that we will use for calculations of the minimal energy states described below.

We choose our units of energy, concentration, and length to reduce the Hamiltonian to the form

$$\mathcal{H} = \frac{1}{2}\sum_{l}\left[(l-l_0)^2 + \frac{R^2\tau}{K}\right]|c_l^m|^2 + \frac{\lambda_3 R}{3!\sqrt{K\lambda_4}}\Upsilon_{m_1,m_2,m_3}^{l_1,l_2,l_3} c_{l_1}^{m_1} c_{l_2}^{m_2} c_{l_3}^{m_3}$$
$$+ \frac{1}{4!}\Upsilon_{m_1,m_2,\overline{m}}^{l_1,l_2,\bar{l}}\Upsilon_{m_3,m_4,-\overline{m}}^{l_3,l_4,\bar{l}} c_{l_1}^{m_1} c_{l_2}^{m_2} c_{l_3}^{m_3} c_{l_4}^{m_4} \tag{6}$$

such that we are left with three dimensionless control parameters: $l_0, \lambda_3 R/\sqrt{K\lambda_4}$, and $R^2\tau/K$. For notational simplicity, we also set $\lambda_3' = \lambda_3 R/\sqrt{K\lambda_4}$.

The ordered (patterned) states are then a linear combination of spherical harmonic basis states described by Eq. (2), where $c_l^m$s are the complex variables that specify the state. The spherical harmonics account for the defects in the pattern induced by the spherical topology, as specified by the Poincaré-Brouwer theorem[30]. We note that because we do not know the precise composition of the primexine, or the effects of the callose wall or any additional chemistry in the space between the cell membrane and the callose wall, the parameters of our model are by necessity phenomenological. However, in principle, with a careful accounting of all the chemistry of the primexine, plasma membrane, and callose wall, it would be possible to independently measure the coefficients described above for a given species and pattern. Next, we describe our method of exploring the phase space of ordered states by finding the set of complex variables, $c_l^m$s, that describe the global minimum energy state.

*Phase Diagram Exploration*

We used simulated annealing (SA) and gradient descent (GD) methods as outlined in *Numerical Recipes*[31] to solve for the minimum energy states of the Hamiltonian in Eq. (6). For simplicity and analytic tractability, we used a single-mode approximation in which we consider patterns at either 1) single $l$ values where $l = l_0$ or 2) the mixing of two adjacent integer values $l$ and $l + 1$ for intermediate values of $l_0$ between $l$ and $l + 1$. We also make some comments on the more general case where we consider the dynamics of the pattern formation.

Since this free energy may potentially have many local minima for a single set of parameters, we used SA to find the global minimum energy state. In this search algorithm, a Metropolis criterion is used in which lower energy states in the phase space are always accepted, while higher energy states are accepted with a Boltzmann probability distribution, $P \propto e^{-\Delta E/T}$, given a temperature-like parameter $T$. The parameter $T$ was tuned to allow the system to escape local minima. Initially, $T$ was chosen to be large enough to allow for an exploration of the whole phase space; $T$ was then lowered with a particular annealing schedule such that the system settled into its global minimum as $T$ became small[31].

To find the appropriate annealing schedule and number of iterations per temperature value, we ran SA enough times to find consistent minimum function values at a given set of parameter values for $l_0$ and $\lambda_3'$. We found that an optimal annealing run started with $T = 1$ and an initial temperature step of $\Delta T = 0.1$. Once we reached a temperature of $T = \Delta T$, we decreased our step size $\Delta T$ by a factor of 10. We continued decreasing the temperature in these incrementally smaller amounts until the observed pattern no longer changed appreciably with further annealing.

We also used GD to ensure that the SA reliably located the global energy minimum for a given parameter set and to test for the presence of local minima. GD is an algorithm that minimizes functions by iteratively moving in the negative direction of the function's gradient until a point with a gradient of zero is found. We were able to calculate the gradient analytically for our model, giving us a substantial computational speed increase.

To confirm the stability of the global minima found via both SA and GD, we diagonalized the Hessian (matrix of second derivatives) and confirmed that all eigenvalues are positive, with the exception of three zero eigenvalues corresponding to the rotations of the sphere. We then comprehensively explored the phase space using both SA and GD by systematically changing the parameter values, $l_0$ and $\lambda'_3$, and recording the effects of those changes to the pattern on the sphere surface. We set $R^2\tau/K = -1$ in this exploration of the phase space to remain in the ordered state, since increasing $R^2\tau/K$ would induce a transition to the unpatterned state.

To study the dynamics of our model, we supposed that the total volume of the primexine condensed and dilute phases are fixed. Therefore, we would generally expect to find a conserved dynamics for our energy. Such a dynamics, consistent with the idea that the free energy is minimized by a spatial modulation with a characteristic wave number $q_0 \equiv 2\pi/\lambda$, is given by

$$\partial_t \psi(\mathbf{x}, t) = D\nabla^2 \frac{\delta \mathcal{H}}{\delta \psi} = D\nabla^2 \left[ K(\nabla^2 + q_0^2)^2 \psi + \tau\psi + \frac{\lambda_3}{2}\psi^2 + \frac{\lambda_4}{6}\psi^3 \right] \quad (7)$$

where we have slightly modified the gradient term in order to more easily integrate the equation of motion. This particular equation of motion is also called the phase-field crystal model[32]. We integrated Eq. (7) using the FiPy package[33], a finite volume solver. Unlike our spherical harmonic method described above, this technique discretizes the sphere and does not preserve rotational symmetry. In addition, we made the wavelength selection weak (i.e, allowed more states away from the characteristic wavelength to contribute to the final pattern) by evolving with $\tau, \lambda_{3,4} \gg K$. For a 2D flat geometry, foamy states are expected[34], and we expect a similar phenomenology on the sphere.

*Evolutionary trait reconstruction*

To examine whether any physical features described by our model of pollen have undergone evolutionary selection, we performed an evolutionary trait reconstruction and subsequent analysis for relative rates of evolution between pattern types across spermatophytes (seed-bearing plants). We first constructed a morphological data set for pollen surface patterns of 2,641 species representing 203 families using the palynology database PalDat[35]. To define a tractable dataset, we limited our morphological analysis to pollen monads, though our theory is potentially general enough to describe any cells of spherical topology.

We restricted our analysis to patterns whose development is commensurate with the underlying assumptions of our model; in order to include a species, we required positive documentation that during the tetrad stage, a given species exhibits plasma membrane undulations with the same wavelength as the mature surface pattern. These data were gathered in a comprehensive review of pollen development literature (see supplemental references). We excluded from our analysis any surface features that demonstrably arise after the dissolution of the callose wall (for example, most echinate spines are derived from tapetal fatty acid deposition)[36]. In these cases, we ignore the post-callose-wall features and analyze the pollen grain as if it did not have them.

We first separated all relevant PalDat SEM images into one of two categories: final pattern is an equilibrium state (i.e., the observed pattern corresponded to an energy minimum from our theory) and final pattern is not at an equilibrium state (i.e., the observed pattern did not correspond to an energy minimum calculated from our theory; instead it was either uniform or foamy). We then measured pollen pattern wavelengths manually in ImageJ. The patterns at equilibrium could be identified as those with surface features with a characteristic (constant) wavelength. All families with patterns in an equilibrium state also had wavelengths >3 μm, except for some species in the family Amaranthaceae, which had wavelengths of 1–3 μm. Conversely, patterns not at equilibrium will not demonstrate a single, constant pattern wavelength but will show a range of wavelengths. Next, we further characterized patterns not at equilibrium into three bins organized by their average pattern wavelength value: $\lambda<1\mu m$, $1<\lambda<3\mu m$, and $\lambda>3\mu m$. Thus, we had four categories to describe pollen from a given family: (1) pattern at equilibrium, $\lambda>3\mu m$ (2) pattern not at equilibrium, $\lambda<1\mu m$, (3) pattern not at equilibrium, $1<\lambda<3\mu m$, (4) pattern not at equilibrium, $\lambda>3\mu m$. Although most equilibrium patterns were formed by exine features, we also considered features previously defined as apertures (i.e., thin regions in the exine) with distinct characteristic wavelengths as equilibrium states. We ignored apertures in a tetrahedral arrangement since these features plausibly result from the geometry of meiosis rather than from the primexine[11]; we analyzed these pollen grains as though the apertures were absent. The observed states not at equilibrium were often foamy (reticulate) with a range of wavelengths. The smallest wavelength category ($\lambda<1$ μm) includes smooth-surfaced pollen.

We used a time-calibrated family-level phylogenetic tree of spermatophytes[37] identified in the integrated Tree of Life (iToL) database[38] to estimate the evolutionary history of these pollen pattern categories. We assigned states to the terminal nodes representing spermatophyte families according to the pattern categories described above; the number of states present in a single family ranged from one to the maximum of four. The Nexus file describing this tree and a fully detailed tree figure are available in the supplemental data.

We used ancestral reconstruction, as implemented in BayesTraits[39] to study the character evolution of patterned states. We used a maximum likelihood algorithm and the multistate model of evolution[40]. We first tested the hypothesis that there is directional evolution either to or from pollen patterns at equilibrium to those that are not at equilibrium. To do this, we defined state A to be category (1), or "at equilibrium," and state B to be categories (2)–(4), or "not at equilibrium". This model is called the "2-state equilibrium model".

We also tested whether there was directional selection for larger pattern wavelengths and therefore more distinctly patterned, polygonal pollen over evolutionary time. For this test, we defined three states (C, D, and E), one for each of the three wavelength categories described above. State C included categories (1) and (2) for all patterns with $\lambda>3\mu m$. State D included all patterns in category (3) not at equilibrium patterns, $1<\lambda<3\mu m$. State E included all patterns in category (4) not at equilibrium patterns, λ<1μm. This model is called the "3-state wavelength model".

## Results

*Microscopy*

We divided the developmental trajectory of pollen in the tetrad state into six distinct stages. In the first stage, after meiosis but before primexine secretion, the plasma membrane did not undulate, there was little or no extracellular material present, and the cell surface was smooth over length scales of about a micron (Fig. 2, col. 1). In the second stage, we observed the primexine material appear on the cell surface (Fig. 2, col. 2, arrowhead). This material was initially uniform in electron density, and the plasma membrane underneath became more irregular, apparently in response to the presence of the material on the cell surface, but there was not a characteristic wavelength in the membrane; the SEM of this developmental stage shows the appearance of a dough-like material on the surface of the cell (Fig. 2, col. 2). In the third stage, the primexine began developing heterogeneities in electron density, and the corresponding SEM showed clumping of the surface material into regions of ~0.5 μm in width, but there was still no characteristic wavelength in the membrane undulation (Fig. 2, col. 3). In the fourth stage, the primexine heterogeneities became more pronounced and the plasma membrane began to undulate with a characteristic wavelength; the SEM at this stage shows distinct domains of separated primexine material on the cell surface with regions of positive curvature separating these domains (Fig. 2, col. 4).

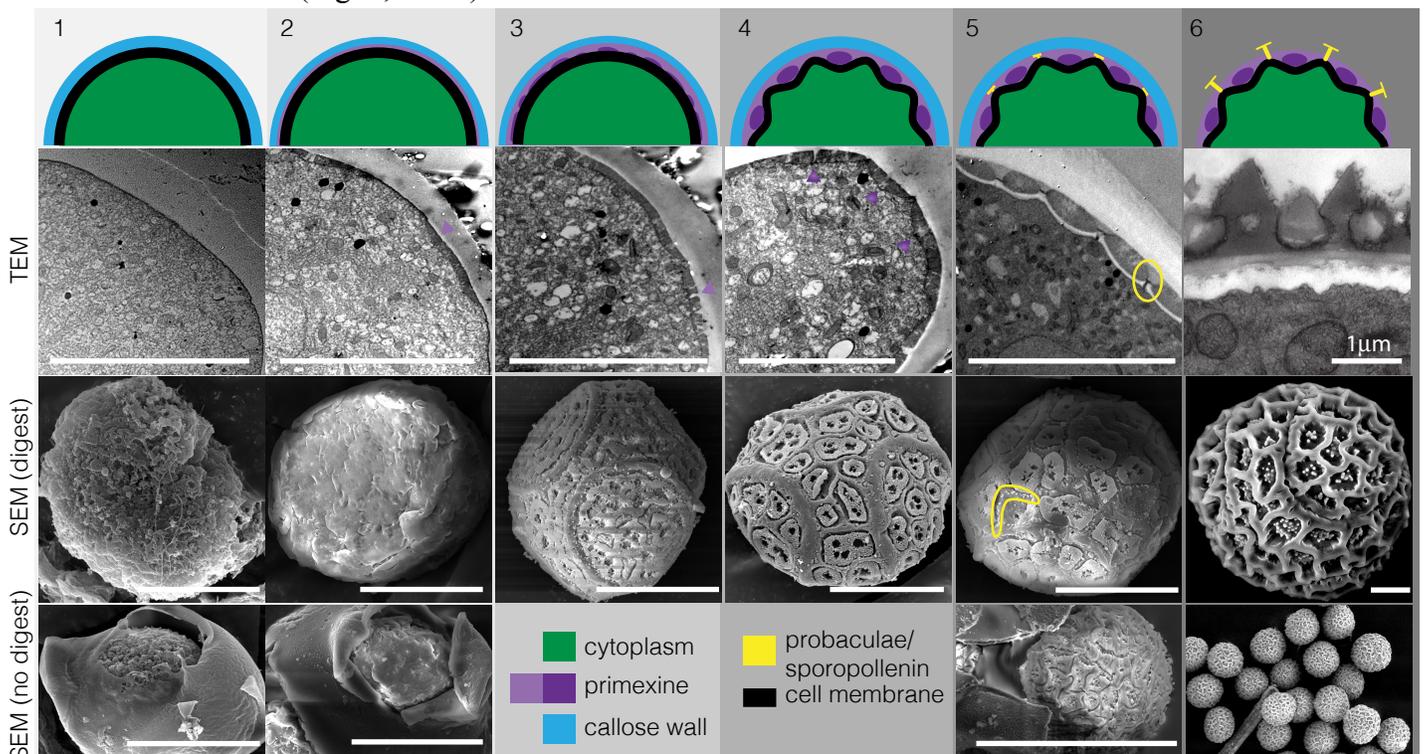

**Figure 2:** *Passiflora incarnata* primexine phase separation. We define five developmental steps of pattern formation occurring after meiosis and prior to callose wall dissolution; the sixth step represents mature pollen. Development proceeds left to right. The first row contains a schematic representation of each step. The second row shows TEM images, the third row shows SEM images with the callose wall enzymatically removed, and the fourth row shows SEM images where the callose wall was mechanically opened but not enzymatically removed. In general, the surface of developing pollen is similar whether the callose wall was removed enzymatically or mechanically. Arrowheads in column 2 indicate the location of the primexine on the cell membrane surface. Arrowheads in columns 4 and 5 indicate the location of dense primexine that causes the cell membrane to locally curve. The circle in column 5 highlights initial formation of probacula/sites of sporopollenin deposition. Column 6 TEM is of mature *Salix alba*[35]. All not labeled scale bars represent 10 *μm*.

In the fifth stage, the phase separation of primexine was complete, with two geometrically regular materials of distinctly different density in contact with the cell membrane. Electron-dense domains (condensed phase) were located on top of regions of negative membrane curvature, and were surrounded by a less electron-dense phase (dilute phase) associated with regions of positive membrane curvature (Fig. 2, col. 5). After primexine phase separation was completed, probacula (sites of sporopollenin accumulation) began forming on the plasma membrane, between electron-dense regions of primexine material and on regions of positive membrane curvature (Fig. 2, col. 5, circled). A dilute phase of primexine can also be observed between the pools of the denser phase in an image of tetrad pollen with a broken callose wall but no enzymatic digestion (Fig. 2, col. 5). The final, sixth stage shows the mature pollen grain with the exine fully deposited onto the patterned primexine; the final exine pattern is formed from the template of low-density primexine material formed during phase separation (Fig. 2, col. 6). While the cytoskeleton is visible in regions of our TEM images, there was no apparent spatial correlation between the location or organization of cytoskeletal elements and the development of membrane undulations, or to the final observed pollen pattern.

*Primexine composition*

The glycosyl composition and linkage analysis of primexine material prepared from developing *Passiflora incarnata* pollen showed a polysaccharide material formed from linkages of a complex mixture of monosaccharides. Given the small amount of material (112.2 μg) we were able to isolate, it was not possible to characterize in detail the chemical structure of the original primexine material. Signal to noise in this analysis was further degraded due to the fact that whole cells were analyzed, such that ~95% of the total residues present were unlinked glucose monomers, and therefore very likely from the cytoplasmic energy stores, not the extracellular matrix. The remaining 5% of residues represented a wide variety of monosaccharides. Several residues, notably galactose (Gal) and mannose (Man), were linked at multiple sites within the monosaccharide, suggesting that the parent material was highly branched. Therefore, after normalizing for glucose content, the constituent monosaccharides and their linkages present during pollen pattern formation were broadly consistent with a mixture of highly branched cellulose, pectin, and hemicellulose-like polymers. The full analysis is available in the supplemental data.

*Phase Diagram Exploration*

To better understand the landscape of patterns generated by this physical mechanism, we explored the equilibrium phase space of the effective Hamiltonian in Eq. (6) by finding the minimum energy states for a range of parameter values. A rich pattern space resulted just from tuning the two dimensionless parameters $l_0$ and $\lambda_3'$ and setting $R^2\tau/K = -1$. Much of this phase space was comprised of patterns with spikes and holes in various polyhedral arrangements; several examples are shown in Figure 3. We found that these patterns could often be categorized into one of three general symmetric types: regular and modified polyhedral spikes; their inverses (duals), in which the spikes become holes; and chiral stripes (Fig. 3). Chiral stripes were only observed when $\lambda_3' = 0$ and $l_0$ was a half integer value (consistent with observations by Sigrist and Matthews)[41]. Chiral stripes have parity symmetry with two chiralities that are energetically degenerate; this degeneracy may be broken by higher-order chiral terms as shown by Dharmavaram and colleagues, thereby biasing a single chirality[17]. These higher-order terms may also plausibly generate the more straight stripes observed in the pollen grains. When this categorization of simple polyhedra or chiral stripes did not apply, the pattern typically represented a mixture of two simpler polyhedral types and/or chiral stripes. Note that for $l = l_0$ states with odd $l_0$, the Gaunt coefficient in front of $\lambda_3'$ vanishes, so the pattern has no $\lambda_3'$ dependence in that case. For even values of $l_0$, the sign of $\lambda_3'$ determined whether the pattern consisted of spikes or holes. At $\lambda_3' = 0$, the spike and hole patterns are degenerate due to the $\psi \to -\psi$ symmetry in the energy. We found that in some regions, the phase space had boundaries across which discontinuous pattern changes were observed (solid lines in Fig. 3). In other regions, patterns gradually changed with systematic tuning of parameters (Fig. 3). We note that we were interested in the broad features of the phase diagram, not the specific characteristics of the phase transitions between patterned states, such as how their continuous or discontinuous nature might change if we include, for example, thermal fluctuations[14,42] or contributions from modes away from $l = l_0$. Finally, we note that in our analysis we found that local minimum states for a given parameter set could match the global minimum state for a separate parameter set (corresponding to areas of coexistence). The occurrence and complexity of these global and local minima is in marked contrast to the planar geometry, where just three stable patterns are observed regardless of the pattern wavelength: uniform stripes, hexagons, or inverted hexagons[43]. Our results are intuitive because on a sphere, none of these three planar patterns can fully wrap the sphere without introducing defects (e.g, pentagonal arrangements of holes and spikes, or points where the stripes collide or end); the many possibilities for accommodating defects yield more possibilities for producing minima in the free energy, as observed in the complexity we find in our phase diagram.

    In studying the dynamics of our model, we found that conserved dynamics indeed yield foamy structures at finite times, as expected from the flat 2d geometry case[34] (Fig. 1, red box). These structures are not identical when different initial conditions are used, so we would generally expect a range of disordered structures in pollen grains of a given non-equilibrating species. We corroborate this prediction with a field of pollen from a single species (*Passiflora incarnata*), which demonstrates that different foamy pollen grains of the same species are slightly different, with a distribution of similar wavelengths comprising the overall reticulate pattern (Fig. 2, col. 5).

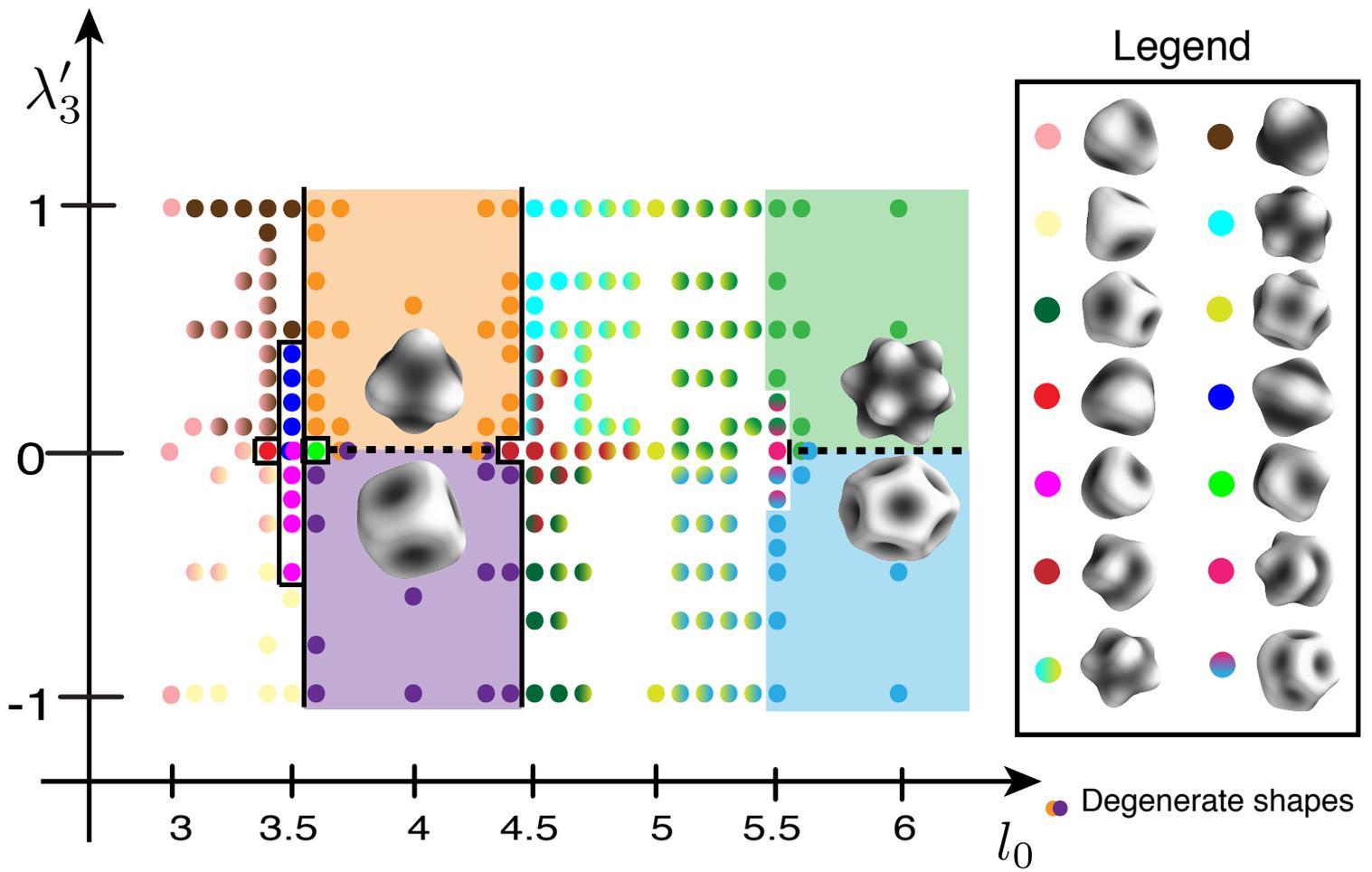

**Figure 3: Phase Diagram of Simulations at Equilibrium.** Calculated energy minima in the ($l_0$, $\lambda_3'$) plane at equilibrium (Eq. 6). Each calculated point is color-coded according to the geometry of the minimum energy state found at that point in the space. Chiral stripe geometry is found at equilibrium when $l_0$ is a half integer and $\lambda_3'=0$. The rest of the space contains polyhedral spike patterns and their inverses. The boundaries between distinct pattern geometries are indicated by black lines. For example, there is a boundary line between $l_0=3.5$ and $l_0=3.6$ from $\lambda_3'=0.1$ to $\lambda_3'=1.0$ across which we observe a discontinuous change in the pattern formed at equilibrium. In contrast, in other regions of the diagram, such as between $l_0=4.5$ and $l_0=5.5$ for $\lambda_3'=1.0$, there is a gradual transition in geometry from one minimum energy state to the next without a distinct boundary line. The legend shows the geometric patterns that correspond to a given color in the phase space. Overlapping dots represent degenerate states. Dots with a gradient of two colors represent intermediate states that are mixtures between two states. Colored shading represents large regions of the space with a single symmetrical pattern.

*Evolutionary Trait Reconstruction*

       We matched patterns generated by our theory to those observed in a pollen database; when we restricted our analysis to monads with documented membrane undulation during development, our dataset represented ~45% of the 453 described families in Sporophyta. This is a minimum set of families potentially described by our theory, since not all families have described pollen and our theory also likely applies to non-monad pollen. This analysis showed that only 27 of 202 included families contain species whose pollen patterns are consistent with an equilibrium state (Fig. 4). Only seven of those 27 families contain species with pollen patterns solely in equilibrium states. The remaining 175 families consist of species exhibiting only non-equilibrated patterns. We found that equilibrium patterns are present throughout angiosperms,

including in gymnosperms, monocots, and eudicots. Notably, equilibrium patterns were absent from the Magnoliids and five other basal families with intermediate branch order between gymnosperms and angiosperms. In gymnosperms, only Welwitschiaceae and Ephedraceae had species with equilibrium pattern states, and both patterns were striped. In monocots, Araceae and Iridaceae had some species with equilibrium patterns, consisting of stripes and polyhedral tiling, respectively. All species in the family Alismataceae had an equilibrium pattern with a polyhedral distribution of pore-like apertures. The rest of the families with some equilibrium states were found in eudicots; their surface patterns were **stripes** (Rubiaceae, Boraginaceae, Scrophulariaceae, Sarraceniaceae, Primulaceae, Lentibulariaceae, Polygalaceae, Acanthaceae, Berberidaceae), **polyhedral spikes** (Asteraceae, Zygophyllaceae, Amaranthaceae, Cucurbitaceae, Alismataceae, Cactaceae, Convolvulaceae, Caryophyllaceae, Polygonaceae, Buxaceae, Polemoniaceae, Martyniaceae, Euphorbiaceae), and **polyhedral holes** (Polemoniaceae, Buxaceae, Polygonaceae, Convolvulaceae, Nyctaginaceae, Zygophyllaceae). Some families had both polyhedral spike and polyhedral hole patterns because the polyhedral arrangement of their apertures fit into a larger exine pattern (see Fig. 1, Convolvulaceae). Of these, only four families contained species with only equilibrated patterns: Polygalaceae, Amaranthaceae, Nyctaginaceae, and Martyniaceae. Examples of each of the pattern types can be found in Figure 1 and the supplemental information.

**Table 1: Model Rates and Probabilities**

| Model | No. rates | $-\ln L$ | Transition rates | Probability of root state | States |
|---|---|---|---|---|---|
| 2-state eq. model | 2 | 31.850555 | qAB = 94.0 qBA = 4.48 | P(A)=0.500 P(B)=0.500 | A: at eq. B: not at eq. |
| Null model for eq. model | 1 | 34.454052 | qAB=qBA =0.235 | P(A)=0.997 P(B)=0.00285 | |
| 3-state λ model | 6 | 125.18975 | qED=29.7 qEC=0 qDE=68.1 qDC=23.2, qCE=58.8, qCD=0 | P(C)=P(D) =P(E)=0.333 | C: λ > 3μm D: 1> λ >3μm E: λ < 1μm |
| Coarser/finer λ model | 2 | 125.39324 | qED=qEC=qDC=22.2, qDE=qCE=qCD=100 | P(C)=P(D) =P(E)=0.333 | |
| Null model for λ model | 1 | 137.01527 | qED=qEC=qDC=qDE=qCE=qCD=0.942 | P(C)=0.135, P(D)=0.111, P(E)=0.754 | |

**Table 2: Hypothesis Tests**

| Models compared | Likelihood ratio | DOF | p-value | Kept model |
|---|---|---|---|---|
| 2-state eq. vs null | 5.21 | 1 | 0.05-0.01 | 2-state eq. |
| 3-state λ vs null | 23.7 | 5 | <<0.01 | 3-state λ |
| 3-state λ vs coarser/finer | 0.407 | 4 | 0.99-0.95 | coarser/finer |

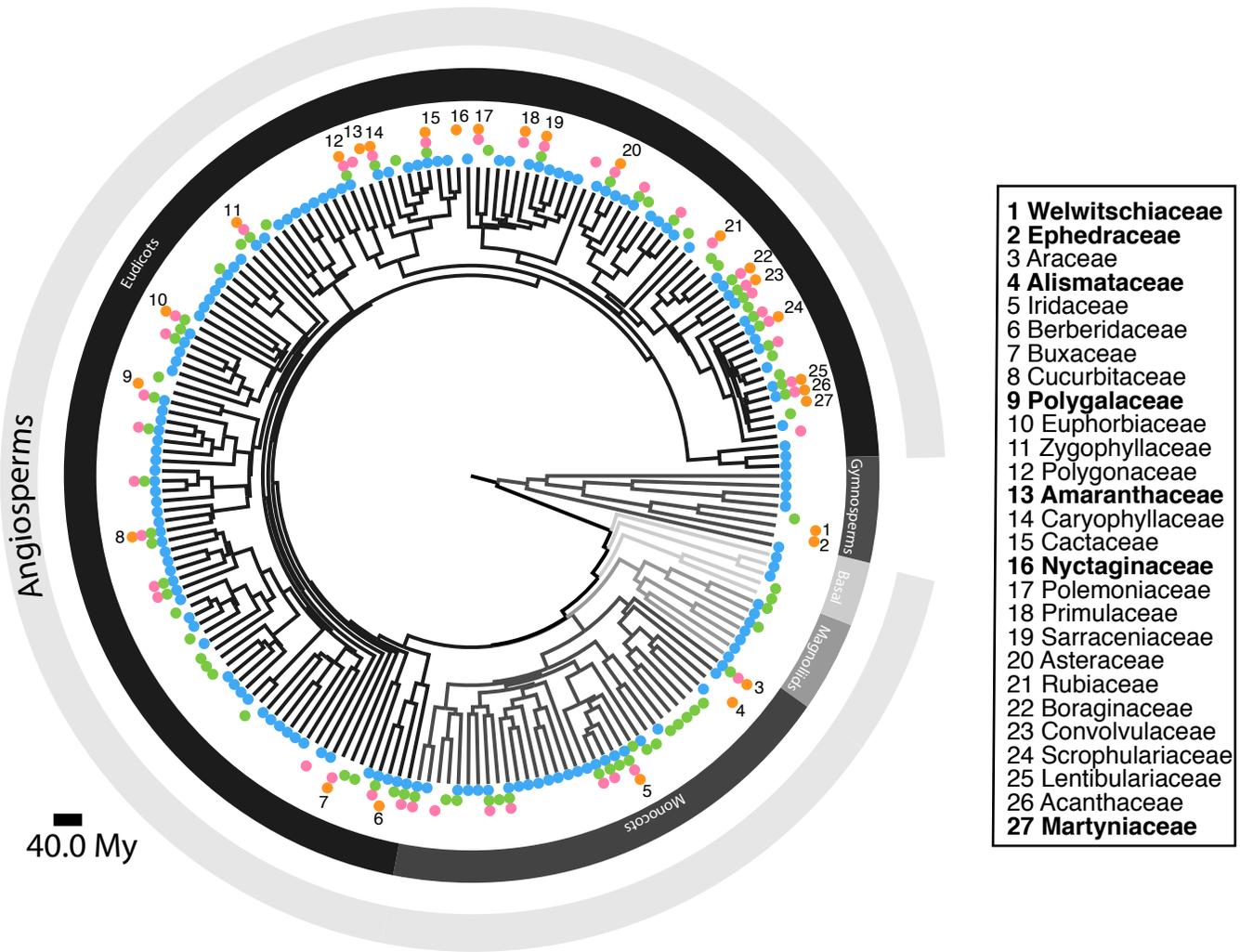

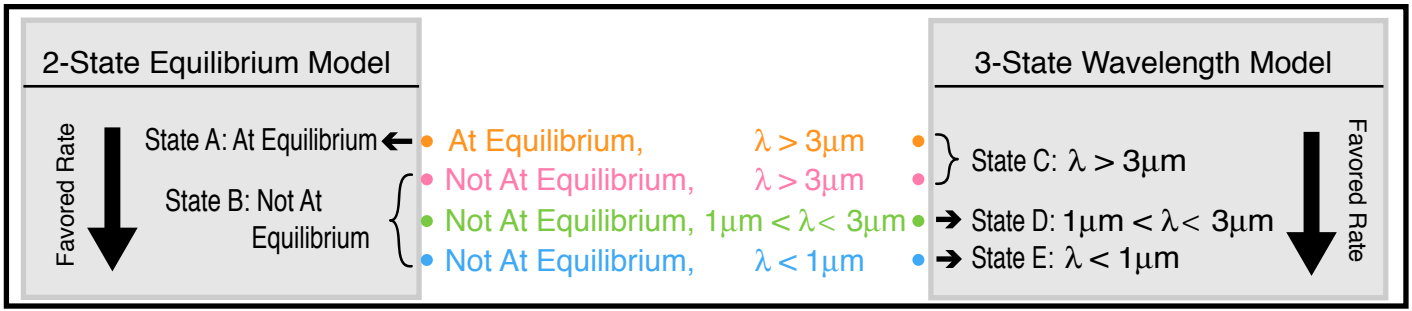

**Figure 4: Angiosperm Phylogenetic Tree with Character States. Top panel:** Phylogenetic tree of spermatophytes with 202 families at terminal taxa. Colored dot represent the character states of the species within each family. Each terminal taxon is labeled with up to four states. The numbered families (27 in total) are those that have species that are in an equilibrium states. The families listed in black have more than one state; families listed in bold (seven of the 27) only have species that are in an equilibrium state. **Bottom panel:** legend for tree and description of categorization of states for two evolutionary models tested. We find that the favored rates are towards not at equilibrium patterns and smaller wavelengths. The scale bar represents 40.0 million years.

We initially hypothesized that if different pollen patterns served different ecophysiological functions, evolution would select for patterns that reach equilibrium during development, since this is presumably a more developmentally predictable and replicable state. We tested this hypothesis using two models of ancestral state reconstruction: a 2-state equilibrium model and a 3-state wavelength model. In the 2-state equilibrium model, we binned our four identified pattern categories of wavelength and equilibrium (see methods) into two evolutionary states, A and B, such that state A is all patterns at equilibrium and state B is all patterns not at equilibrium (Fig. 4, bottom panel). The log-likelihood ratio of the 2-state equilibrium model compared to the null model (where both rates are equal) was 5.21, so with one degree of freedom, the p-value was between 0.01 and 0.05 (Table 2). We therefore reject the null model and find that the rate of evolutionary transition from equilibrium to non-equilibrium patterns is ~20-fold greater than the reverse rate (Table 1, qAB=94.0, qBA = 4.48). We also found that the state at the root of spermatophytes had equal probability of being at equilibrium or non-equilibrium.

We then tested the 3-state wavelength model by re-sorting categories (1)–(4) so that state C represented all patterns with wavelengths greater than 3 µm, state D represented patterns with wavelengths between 1 and 3 µm, and state E represented patterns with wavelengths less than 1 µm (see methods and Fig. 4, bottom panel). We first compared the 3-state wavelength model to the null model and found a likelihood ratio of 23.7 given five degrees of freedom, for a p-value <<0.01. Therefore, we reject the null hypothesis and accept the 3-state wavelength model. We next compared the 3-state wavelength model with a simpler coarser/finer model where we restricted all rates towards larger wavelengths (coarser) to be equal to each other (Table 1, qED=qEC=qDC) and all rates towards smaller wavelengths (finer) to be equal to each other (Table 1, qDE=qCE=qCD). The likelihood ratio between these two models resulted in a p-value between 0.99 and 0.95, such that there was no significant difference between them. It is therefore likely to be the case that pollen evolves more rapidly from equilibrated polygonal patterns to finely reticulated or bumpy patterns than the reverse, and that any more complicated model of pattern type evolution will be over fit. In other words, evolution seems to favor pollen that never reaches equilibrated patterns, and similarly, foamy (reticulate) or unpatterned pollen seems favored over the more interesting-to-humans pollen with well-defined polygonal patterns.

**Discussion**

We observed that both the electron density and the surface distribution of the primexine of *Passiflora incarnata* change, becoming inhomogeneous, during pattern development. Because the primexine electron density is initially uniform but subsequently separates into two distinct electron densities, primexine development is consistent with a phase separation into a dense and a dilute phase. The phase transition of polysaccharide materials of this kind is expected in the absence of cross-linking factors (perhaps, for example, into phases with more- and less-branched polymers). Additionally, we observed that the denser phase correlates to the plasma membrane undulations (with discrete patches of dense material sitting inside the dips in the membrane).

Therefore, our data suggest that the more dense primexine regions cause the plasma membrane to curve away from its initially featureless, spherical shape. The final pollen exine pattern is then negatively templated by the pooled dense primexine and correlated membrane curvature.

A previous study of *Brassica campestris* pollen, another reticulate species, also demonstrated the same deposition of primexine on the plasma membrane surface[44] followed by plasma membrane undulations correlated to a dense primexine phase. In addition to the many reticulate species whose patterns seem to be templated by plasma membrane undulations, species with other surface patterns such as the polygonal holes of *Ipomoea purpureae*[45] or the polygonal spikes of *Farfugium japonicum*[46] also exhibit early membrane undulations at the same wavelength as the mature pattern features. However, primexine was not preserved in these studies[23].

Although we were unable to determine the exact chemical composition of the primexine, the constituent monosaccharides and their linkages are consistent with a mixture of cellulose, pectin- and hemicellulose-like polysaccharides. Mixtures of different polysaccharides tend to phase separate unless a cross-linker actively prevents them from demixing[26], such that phase separation of primexine material on the surface of a developing pollen cell is perhaps not surprising.

Our theory shows that this phase separation of a material on the surface of a spherical cell, when coupled to membrane elasticity (i.e., membrane buckling), yields an effective free energy that exhibits spatially modulated phases. This effective free energy, using both single-mode and two-mode approximations, produced equilibrium states corresponding to a variety of spikes, holes, and chiral stripes on the surface of a sphere. These equilibrium patterns generated by our theory also correspond to about ten percent of the pollen patterns documented in PalDat. We expect that other highly ordered, patterned pollen may also fit our model when we include more modes.

The more disorganized patterns observed in ~90% of analyzed species may be explained by the dynamics of the process encoded by our model. Indeed, if we arrest the dynamics after some short time (before equilibrium can occur), we find states that resemble the foamy, more disordered pollen structures. In the planar case, some of these foamy structures may even be relatively stable, as discussed in more detail by Guttenberg and colleagues[34]. Applying the techniques in this work to the surface of a sphere would be an interesting topic for future research.

Given the observation of so many species that either have a non-equilibrated pattern, or no pattern at all, it is worth thinking about what this means in the context of our general physical theory. One possibility is that most plant materials have effective free energy parameters that barely favor phase separation of the primexine. This possibility would explain both the repeated evolution of featureless pollen and the high abundance of disordered structures, both of which could result from the slower kinetics and enhanced fluctuations that one would generally expect near a phase transition, especially if the phase transition has only a weakly discontinuous character. In mixtures that start near such a critical point, small variations of the parameters (induced, for example, by small changes in chemical composition of the primexine) could induce large changes in the patterning. This possibility might lend additional weight to our physical theory as an explanation of the observed pattern diversity; small evolutionary shifts in primexine composition could fundamentally alter the mature pollen pattern, leading to the relatively large shifts in pollen patterns in short periods of time that have demonstrably occurred in evolution. Although we did not detect an elevated rate of appearance of equilibrium patterns, the tree is consistent with many instances of equilibrium pollen patterns arising from evidently non-equilibrium patterns of recent ancestors. For example, the families Asteraceae, Sarraceniaceae, and Cactaceae all exhibit equilibrated patterns that are nested in clades in which the other families exhibit only non-equilibrated patterns. Another possible explanation for the prevalence of the disordered states is that primexine phase separation is typically arrested by sporopollenin deposition before it can bring the pollen grain into an equilibrium pattern. In addition, cross-linkers such as calcium ions are often found in plant cell walls, the presence of which might also contribute to the formation of the more disordered patterns by arresting the underlying separation dynamics.

After classifying extant pollen patterns as either equilibrium states versus kinetically arrested or generally disordered patterns, both of which are predicted by the physical mechanism proposed here, we conclude from an evolutionary analysis that the highly ordered patterns for which pollen are famous have not arisen under strong selection. In fact, our results are more consistent with an evolutionary bias toward unpatterned, typically foamy (reticulate) states. This evolutionary result is also consistent with our physical picture, since the constituents of the primexine are naturally phase-separating compounds and should induce the patterning without any additional biological control. So, perhaps the exine patterns that give pollen their fascinating variety do not serve any particular purpose, but are rather a natural consequence of the composition of the primexine and simple physical principles.

There is much room for future work. Our theory, and its apparent reification in pollen development, describe a novel and robust mechanism for repeatedly patterning surfaces at both micron and nanometer scales. Therefore, it would be of basic interest to materials science to understand how to program the general parameters of our theory in polymer chemistry. By fully characterizing the primexine material, it would be possible to study its phase properties and their contribution to the pattern-governing parameters in our model. Finally, in contrast to the currently employed pollen descriptive scheme of overlapping categories of unit, polarity, aperture, ornamentation, and wall structure, nearly all unique pollen patterns can be fully recapitulated by a unique set of parameters in our Hamiltonian (eq. 6). It may be useful in the future to describe pollen species by these unique energetic parameters; this scheme also has the advantage that these energetic parameters will ultimately map to the biochemistry and timing of pollen development.

**Acknowledgements**
The authors gratefully acknowledge the use of the Electron Microscopy Resource Laboratory at the Perelman School of Medicine at the University of Pennsylvania. This work was supported by the Chemical Sciences, Geosciences and Biosciences Division, Office of Basic Energy Sciences, U.S. Department of Energy grant (DE-SC0015662) to Parastoo Azadi at the Complex Carbohydrate Research Center. This work was also supported by a Kaufman Foundation New Initiative Award to A.M.S., a Packard Foundation Fellowship to A.M.S., and NSF-1351935 to A.M.S.; we also gratefully acknowledge the help of summer interns through the PennLENS program supported by NSF-1351935. E.M.H. and M.O.L. were supported in part by a Simons Investigator Grant to Randall D. Kamien. M.O.L. gratefully acknowledges partial funding from the Neutron Sciences Directorate (Oak Ridge National Laboratory), sponsored by the U.S. Department of Energy, Office of Basic Energy Sciences.**References**

1. Blackmore, S., Wortley, A.H., Skvarla, J.J., & Rowley, J.R. Pollen wall development in flowering plants. *New Phytol*. **174**, 483-498 (2007)

2. Locke, M. *Microscopic Anatomy of Invertebrates*. Eds Harrison FW, Locke M (Wiley-Liss, New York) **11A**, 75–138 (1998)

3. Ariizumi, T. & Toriyama, K. Genetic regulation of sporopollenin synthesis and pollen exine development. *Annu Rev Plant Biol*. **62**, 437–460 (2011)

4. Blackmore, S. & Crane, P.R. The evolution of apertures in the spores and pollen grains of embryophytes. *Rep Biol*. 159-182 (1998)

5. Katifori, E., Alben, S., Cerda, E., Nelson, D.R., & Dumais, J. Foldable structures and the natural design of pollen grains. *PNAS* **107**, 7635-7639 (2010)

6. Owen, H.A. & Makaroff, C.A. Ultrastructure of microsporogenesis and microgametogenesis in *Arabidopsis thaliana* (L.) Heynh. ecotype Wassilewskija (Brassicaceae). *Protoplasma*. **185**, 7-21 (1995)

7. Nishikawa, S., Zinkl, G.M., Swanson, R.J., Maruyama, D., & Preuss, D. Callose ($\beta$-1,3 glucan) is essential for *Arabidopsis* pollen wall patterning, but not tube growth. *BMC Plant Biol*. **5**, 22 (2005)

8. Heslop-Harrison, J. Tapetal origins of pollen-coat substances in Lilium. *New Phytol*. **67**, 779-786 (1968)

9. Skvarla, J.J. & Larson, D.A. Fine Structural Studies of Zea mays Pollen I: Cell Membranes and Exine Ontogeny. *Am J Bot*. **53**, 1112-1125 (1966)

10. Godwin, H., Echlin, P., & Chapman, B. The development of the pollen grain wall in *Ipomoea purpurea* (L.) Roth. *Rev. Palaeobot. Palynol*. **3**, 181-195 . (1967)

11. Rowley, J.R. Germinal Apertural Formation in Pollen. *Taxon* **24**, 17-25 (1975)

12. Furness, C.A. A review of spiraperturate pollen. *Pollen et Spores* **27**, 307-320 (1985)20

13. Albert, B., Nadot, S., Dreyer, L., & Ressayre, A. The influence of tetrad shape and intersporal callose wall formation on pollen aperture pattern ontogeny in two eudicots species. *Ann. Bot.* **106**, 557-564 (2010)

14. Lavrentovich, M.O., Horsley, E.M., Radja, A., Sweeney, A.M., & Kamien, R.D. First-order patterning transitions on a sphere as a route to cell morphology. *PNAS*. **113**, 5189-5194 (2016)

15. Leibler, S. & Andelman, D. Ordered and curved meso-structures in membranes and amphiphilic films. *J Phys*. **48**, 2013–2018 (1987)

16. Dharmavaram, S., Xie, F., Klug, W., Rudnick, J., & Bruinsma, R. Landau theory and the emergence of chirality in viral capsids. *EPL* **116**, 26002 (2016)

17. Dharmavaram, S., Xie, F., Klug, W., Rudnick, J., & Bruinsma, R. Orientational phase transitions and the assembly of virus capsids. *Phys. Rev. E*. **95**, 062402 (2017)

18. Andelman, D., Kawakatsu, T., Kawasaki, K. Equilibrium shape of two-component unilamellar membranes and vesicles. *Europhys. Lett.* **19**, 57-62 (1992)

19. Taniguchi, T., Kawasaki, K., Andelman, D. and Kawakatsu, T. Phase transitions and shapes of two component membranes and vesicles II : weak segregation limit. *J. Phys. II France*. 4, 1333-1362 (1994)

20. Zhang, L., Wang, L., & Lin, J. Defect structures and ordering behaviours of diblock copolymers self-assembling on spherical substrates. *Soft Matter*. **10**, 6713–6721 (2014)

21. Sannier, J., Baker, W.J., Anstett, M.C., & Nadot, S. A comparative analysis of pollinator type and pollen ornamentation in the Araceae and the Arecaceae, two unrelated families of the monocots. *BioMed Central*. **2**, 145 (2009)

22. Furness, C.A. & Rudall, P.J. Pollen aperture evolution – a crucial factor for eudicot success? *Trends in Plant Sci*. **9**, 154-158 (2004)

23. Gabarayeva, N.I. & Grigorjeva, V.V. Exine development in *Stangeria eriopus* (Stangeriaceae): ultrastructure and substructure, sporopollenin accumulation, the equivocal character of the aperture, and stereology of microspore organelles. *Rev. Palaeobot. Palynol.* **122**, 185-218 (2002)

24. Kirkpatrick, A.B. & Owen, H.A. Observation of early pollen exine patterning by scanning electron microscopy. *Microsc. Microanal*. **19**, 134-135 (2013)

25. Santander, J. *et al*. Mechanisms of intrinsic resistance to antimicrobial peptides of *Edwardsiella ictaluri* and its influence on fish gut inflammation and virulence. *Microbiology.* **159**, 1471-1486 (2013)

26. Tolstoguzov, V. Phase behavior in mixed polysaccharide systems. *Food Polysaccharides and Their Applications $2^{nd}$ ed.* Eds. Stephen AM, Phillips GO, Williams PA. Boca Raton: Taylor & Francis 589-627 (2006)

27. Kim, S. & Willett, J. L. Isolation of amylose from starch solutions by phase separation. *Starch*. **56**, 29-36 (2004)

28. Abramowitz, M. & Stegun, I.A. *Handbook of Mathematical Functions* (National Bureau of Standards, Washington, DC) (1972)

29. Johansson, H.T. & Forssén, C. Fast and accurate evaluation of Wigner 3j, 6j, and 9j symbols using prime factorisation and multi-word integer arithmetic. ArXiv: 1504.08329 (2015)

30. Kamien, R.D. The geometry of soft materials: A primer. *Rev Mod Phys.* **74**, 953–971. (2002)

31. Press, W.H., Teukolsky, S.A., Vetterling, W.T., & Flannery, B.P. *Numerical Recipes : the Art of Scientific Computing.* Cambridge [Cambridgeshire]: Cambridge University Press (1986)

32. Elder, K.R., Katakowski, M., Haataja, M., & Grant, M. Modeling Elasticity in Crystal Growth. *Phys Rev Lett.* **88**, 245701 (2002)

33. Guyer, J.E., Wheeler, D., & Warren, J.A. FiPy: Partial Differential Equations with Python. *Comput Sci Eng.* **11**, 6-15 (2009)

34. Guttenberg, N., Goldenfeld, N. & Dantzig. J. Emergence of foams from the breakdown of the phase field crystal model. *Phys. Rev. E.* **81**, 065301 (2010)

35. PalDat – a palynological database (2000 onwards, www.paldat.org), downloaded on 9/27/16)

36. Weber, M., Halbritter, H., & Hesse, M. The spiny pollen wall in Sauromatum (Araceae) – with special reference to the endexine. *Int. J. Plant Sci.* **159**,744-749 (1998)

37. Harris, L. W., Jonathan Davies, T. A Complete Fossil-Calibrated Phylogeny of Seed Plant Families as a Tool for Comparative Analyses: Testing the 'Time for Speciation' Hypothesis. *PLoS ONE*. **10**, e0162907 (2016)

38. Hinchliff, C. E. *et al*. Synthesis of phylogeny and taxonomy into a comprehensive tree of life. *PNAS*. **112**, 12764-12769 (2015)

39. Pagel, M., Meade, A., 2006. BayesTraits. Available from: <http://www.evolution.rdg.ac.uk/BayesTraits.html>.

40. Pagel, M. The Maximum likelihood approach to reconstructing ancestral character states of discrete characters on phylogenies. *Syst Biol*. **48**, 612-622 (1999)

41. Sigrist, R., Matthews, P. Symmetric spiral patterns on spheres. *SIAM J Appl Dyn Syst*. **10**, 1177–1211 (2011)

42. Brazovskii, S.A. Phase transition of an isotropic system to a nonuniform state. *J Exp Theor Phys*. **68**,175-185 (1975)

43. Brazovskii, S.A., Dzyaloshinskii, I.E., & Muratov, A.R. Theory of weak crystallization. *J Exp Theor Phys*. **66**, 625-633 (1987)

44. Fitzgerald, M.A. & Knox, R.B. Initiation of primexine in freeze-substituted microscpores of *Brassica campestris*. *Sex plant reprod*. **8**, 99-104 (1995)

45. Takahashi, M. Development of the echinate pollen wall in *Farfugium japonicum* (Compositae: Senecionea). *Bot. Mag*. **102**, 219-234 (1989)

**Supplemental Data**

- Figure S1: Detailed phylogenetic tree with 202 labeled families in spermatophytes. Colored dot represent the character states of the species within each family. Each terminal taxon is labeled with up to four states. The scale bar represents 40.0 million years.
- Table S1: List of papers used to categorize pollen.
- File S1: Glycosyl composition analysis completed at the Complex Carbohydrate Research Center at the University of Georgia for *Passiflora incaranta*.
- File S2: Glycosyl linkages analysis completed at the Complex Carbohydrate Research Center at the University of Georgia for *Passiflora incaranta*.
- File S3: Nexus file used in our analysis. This file includes the morphological data for all spermatophyte families and the phylogenetic tree.

**Figure S1: Phylogenetic Tree with Family Names**

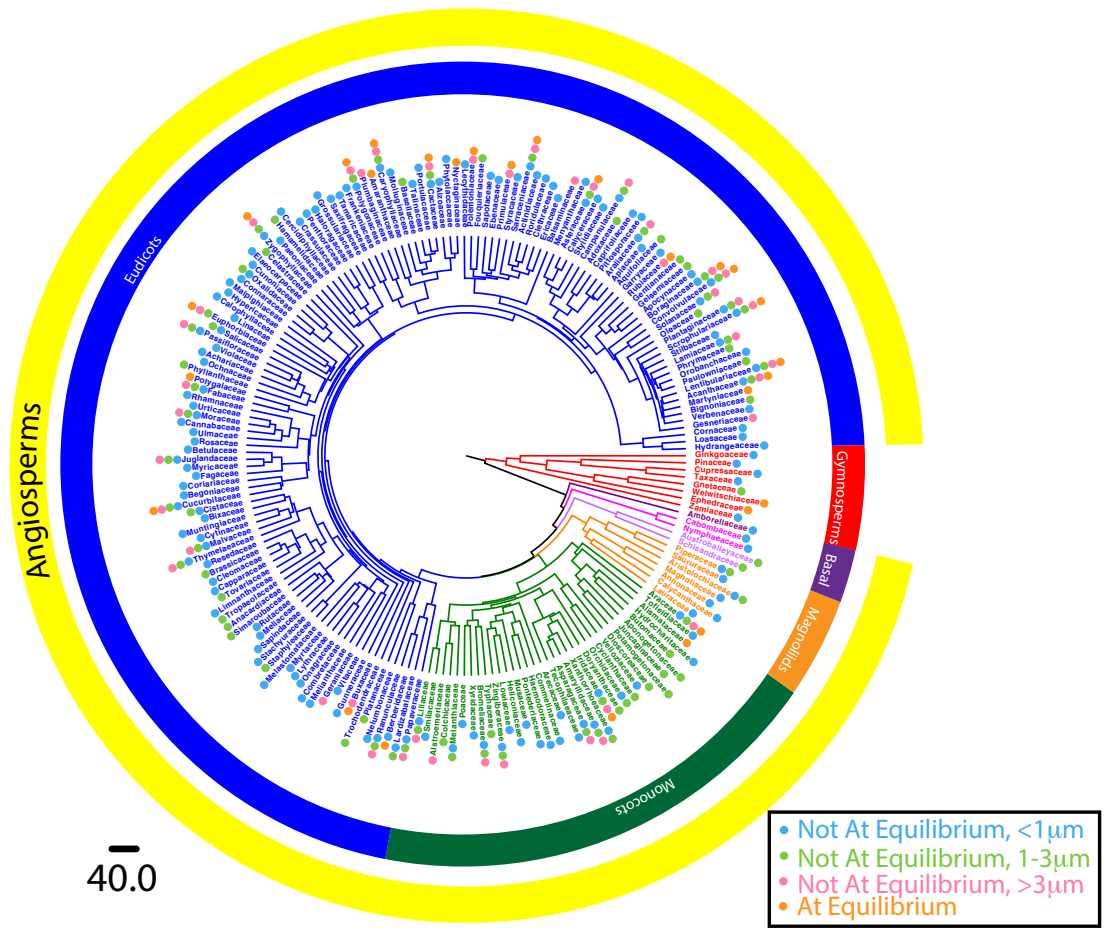

**Supplemental Figure 1: Detailed Angiosperm Phylogenetic Tree with Character States.**
Phylogenetic tree of spermatophytes with 202 families at terminal taxa. Colored dot represent the character states of the species within each family. Each terminal taxon is labeled with the family name and with up to four states. The numbered families (27 in total) are those which have species that are in an equilibrium states. The families listed in black have more than one state; families listed in bold (seven of the 27) only have species that are in an equilibrium state. The scale bar represents 40.0 million years.

# Table S1: List of papers used to categorize pollen

Literature used to categorize pollen species into equilibrium or non-equilibrium states

| Family | References |
|---|---|
| Annonaceae | Gabarayeva 1995 |
| Araceae | Anger & Weber 2006, Weber et al. 1998 |
| Aristolochiaceae | Gonzalez et al 2001, Polevova 2015 |
| Asteraceae | Blackmore & Barnes 1988, Blackmore et al. 2010, Tomb et al. 1974, Takahashi 1989, Dickinson & Potter 1976, Blackmore & Barnes 1987, Horner Jr. & Pearson 1978 |
| Austrobaileyaceae | Zavada 1984 |
| Boraginaceae | Gabarayeva et al. 2011 |
| Brassicaceae | Fitzgerald & Knox 1995 |
| Caryophyllaceae | Heslop-Harrison 1963, Audran & Batcho 1981, Shoup et al 1980 |
| Convolvulaceae | Echlin et al. 1966 |
| Ephedraceae | Doores et al. 2007 |
| Gnetaceae | Yao et al, 2004 |
| Heliconiaceae | Stone 1987 |
| Hydrocharitaceae | Takahashi 1994 |
| Lauraceae | Stone 1987, Rowley & Vasanthy 1993 |
| Liliaceae | Sheldon & Dickinson 1986, Heslop-Harrison 1968 |
| Malvaceae | Takahashi & Kouchi 1988 |
| Nyctaginaceae | Takahashi & Skvarla 1991 |
| Nymphaeaceae | Takahashi 1992 |
| Papaveraceae | Romero et al 2003 |
| Welwitschiaceae | Doores t al. 2007 |

## Literature cited

Gabarayeva, N.I. Pollen wall and tapetum development in *Anaxagorea brevipes* (Annonaceae): sporoderm substructure, cytoskeleton, sporopollenin precursor particles, and the endexine problem. *Rev. Palaebot. Palynol.* **85**, 123-152 (1995)

Anger, E.M. & Weber, M. Pollen-wall formation in *Arum alpinum*. *Ann. Bot.* **97**, 239-244 (2006)

Weber, M., Halbritter, H., & Hesse, M. The spiny pollen wall in Sauromatum (Araceae)-with special reference to the endexine. *Int. J. Plant Sci.* **159**, 744-749 (1998)

González, F., Rudall Fls, P.J., Furness, C.A. Microsporogenesis and systematics of Aristolochiaceae. *Bot. J. Linn. Soc.* **137**, 221-242 (2001)

Polevova, S.V. Ultrastructure and development of sporoderm in *Aristolochia clematitis* (Aristolochiaceae). *Rev. Palaeobot. Palynol.* **222**, 104-115 (2015)

Blackmore, S. & Barnes, S.H. Pollen ontogeny in *Catananche caerulea* L. (Compositae: Lactuceae) I. Premeiotic phase to establishment of tetrads. *Ann. Bot.* **62**, 605-614 (1988)

Blackmore, S., Wortley, A.H., Skvarla, J.J., & Gabarayeva, N.I. Developmental origins of structural diversity in pollen walls of Compositae. *Plant Syst. Evol.* **284**, 17-32 (2010)

Tomb, A.S., Larson, D.A., & Skvarla, J.J. Pollen morphology and detailed structure of family Compositae, tribe Cichorieae. I. Subtribe Stephanomeriinae. *Am. J. Bot.* **61**, 486-498 (1974)

Takahashi, M. Pattern determination of the exine in Caesalpinia japonica (Leguminosae: Caesalpinioideae). *Am. J. Bot.* **76**, 1615-1626 (1989)

Dickinson, H.G., Potter, U. The development of patterning in the alveolar sexine of *Cosmos bipinnatus*. *New Phytol.* **76**, 543-550 (1976)

Blackmore S. & Barnes, S.H. Pollen wall morphogenesis in *Tragopogon porrifolius* L. (Compositae: Lactuceae) and its taxonomic significance. *Rev. Palaeobot. Palynol.* **52**, 233-246 (1987)

Horner Jr., H. T. & Pearson, C. B. Pollen wall and aperture development in Helianthus annuus (Compositae: Heliantheae). *Am. J. Bot.* **65**, 293-309 (1978)

Zavada, M. S. Pollen wall development of Austrobaileya maculata. *Int. J. Plant Sci.* **145**, 11-21 (1984)

Gabarayeva, N., Grigorjeva, V., & Polevova, S. Exine and tapetum development in *Symphytum officinale* (Boraginaceae). Exine substructure and its interpretation. *Plant Syst. Evol.* **296**, 101-120 (2011)

Fitzgerald, M. A. and Knox, R. B. Initiation of primexine in freeze-substituted microspores of *Brassica campestris*. *Sex. Plant Reprod.* **8**, 99-104 (1995)

Heslop-Harrison, J. An ultrastructural study of pollen wall ontogeny in Silene pendula. *Grana*. **4**, 7-24 (1963)

Audran, J., Batcho, M. Microsporogenesis and pollen grains in *Silene dioica* (L.) Cl. and alterations in its anthers parasite by *Ustilago violaceae* (Pers.) Rouss. (Ustilaginales). *Acta. Soc. Bot. Pol.* **50**, 29-32 (1981)

Shoup, J. R., Overton, J. & Ruddat, M. Ultrastructure and development of the sexine in the pollen wall of *Silene alba* (Caryophyllaceae). *Bot. Gaz.* **141**, 379-388 (1980)

Echlin, P., Chapman, B., Godwin, H., & Angold, R. The fine structure and development of the polen of Helleborus foetidus L. *J. Electron Microsc.* **2**, 315-316 (1966)

Doores, A. S., Osborn, J. M., El-Ghazaly, G. Pollen ontogeny in *Ephedra americana* (Gnetales). *Int. J. Plant Sci*. **168**, 985-997 (2007)

Yao, Y. F., Xi, Y. Z., Geng, B. Y., & Li, C. S. The exine ultrastructure of pollen grains in *Gnetum* (Gnetaceae) from China and its bearing on the relationship with the ANITA Group. *Bot. J. Linn. Soc.* **146**, 415-425 (2004)

Stone, D. E. Developmental evidence for the convergence of *Sassafras* (Laurales) and *Heliconia* (Zingiberales) pollen. *Grana*. **26**, 179-191 (1987)

Takahashi, M. Pollen development in a submerged plant, *Ottelia alismoides* (L.) Pers. (Hydrocharitaceae). *J. Plant Res.* **107**, 161-164 (1994)

Rowley, J. R. & Vasanthy, G. Exine development, structure, and resistance in pollen of *Cinnamomum* (Lauraceae). *Grana*. **32**, 49-53 (1993)

Sheldon, J. M. & Dickinson, H. G. Pollen wall formation in *Lilium*: The effect of chaotropic agents, and the organization of the microtubular cytoskeleton during pattern development. *Planta*. **168**, 11-23 (1986)

Heslop-Harrison, J. Tapetal origin of pollen-coat substances in *Lilium*. *New Phytol.* **67**, 779-786 (1968)

Takahashi, M. & Kouchi, J. Ontogenetic development of spinous exine in Hibiscus syriacus (Malvaceae). *Am. J. Bot.* **75**, 1549-1558 (1988)

Takahashi, M., Skvarla, J. J. Exine pattern formation by plasma membrane in Bougainvillea Spectabilis Willd. (Nyctaginaceae). *Am. J. Bot.* **78**, 1063-1069 (1991)

Takahashi, M. Development of spinous exine in *Nuphar japonicum* DeCandolle (Nymphaeaceae). *Rev. Palaeobot. Palynol.* **75**, 317-322 (1992)

Romero, A. T., Salinas M. J., & Fernández, M. C. Pollen wall development in *Hypecoum imberbe* Sm. (Fumariaceae). *Grana*. **42**, 91-101 (2003)

File S1: Glycosyl composition analysis completed at the Complex Carbohydrate Research Center at the University of Georgia for *Passiflora incaranta*.

Date:  2/28/17

Investigator:  **Asja Radja**
University of Pennsalvania
209 S 33rd Street
Philadelphia, PA 19104

Subject:  Glycosyl composition analysis.

Sample  Sample #1

CCRC Code:  AR021617

Analyst:  Ian Black

**Methods:**

**Should any of these data be used in a publication, please include the following statement in the acknowledgment: This work was supported by the Chemical Sciences, Geosciences and Biosciences Division, Office of Basic Energy Sciences, U.S. Department of Energy grant (DE-SC0015662) to Parastoo Azadi " at the Complex Carbohydrate Research Center.**

Glycosyl composition

Glycosyl composition analysis was performed by combined gas chromatography/mass spectrometry (GC/MS) of the per-*O*-trimethylsilyl (TMS) derivatives of the monosaccharide methyl glycosides produced from the sample by acidic methanolysis as described previously by Santander *et al*. (2013) *Microbiology* **159**:1471.

Briefly, the sample (300 ug) was heated with methanolic HCl in a sealed screw-top glass test tube for 17 h at 80 °C. After cooling and removal of the solvent under a stream of nitrogen, the sample was treated with a mixture of methanol, pyridine, and acetic anhydride for 30 min. The solvents were evaporated, and the sample was derivatized with Tri-Sil® (Pierce) at 80 °C for 30 min. GC/MS analysis of the TMS methyl glycosides was performed on an Agilent 7890A GC interfaced to a 5975C MSD, using an Supelco Equity-1 fused silica capillary column (30 m × 0.25 mm I

# Glycosyl Composition Analysis

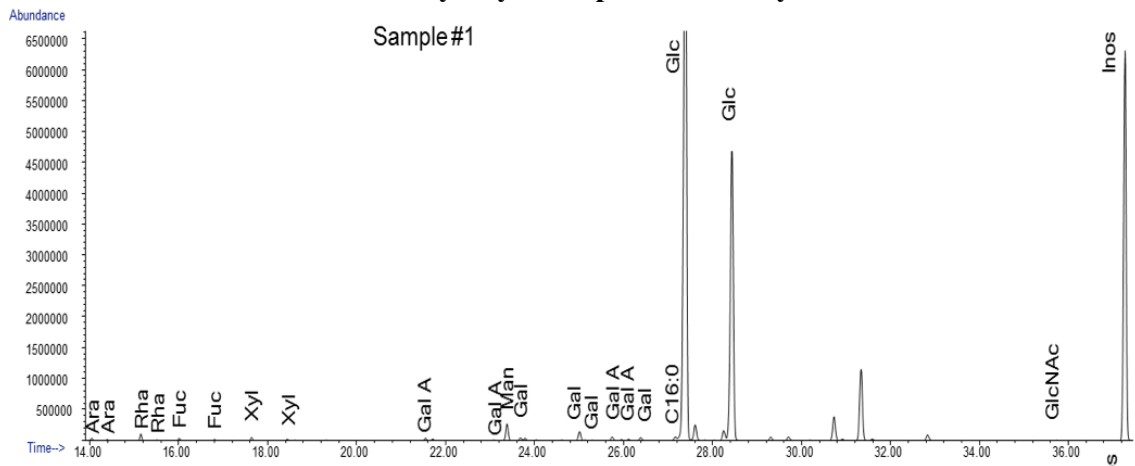

The GC chromatograms of the compositional analysis.

| Sample | Glycosyl residue | Mass (μg) | Mol % |
|---|---|---|---|
| Sample #1 | Ribose (Rib) | n.d. | - |
| | Arabinose (Ara) | 0.2 | 0.2 |
| | Rhamnose (Rha) | 0.7 | 0.7 |
| | Fucose (Fuc) | 0.2 | 0.2 |
| | Xylose (Xyl) | 0.4 | 0.4 |
| | Glucuronic Acid (GlcA) | n.d. | - |
| | Galacturonic acid (GalA) | 0.8 | 0.6 |
| | Mannose (Man) | 1.5 | 1.3 |
| | Galactose (Gal) | 1.4 | 1.2 |
| | Glucose (Glc) | 106.9 | 95.2 |
| | N-Acetyl Galactosamine (GalNAc) | n.d. | - |
| | N-Acetyl Glucosamine (GlcNAc) | 0.1 | 0.1 |
| | N-Acetyl Manosamine (ManNAc) | n.d. | - |
| | | 112.2 | |

The estimated amounts and mole percentage of each detected monosaccharide in the sample.

**File S2: Glycosyl linkages analysis completed at the Complex Carbohydrate Research Center at the University of Georgia for *Passiflora incaranta*.**

| | |
|---|---|
| **Date:** | 3/6/17 |
| **Investigator:** | **Asja Radja** |
| | University of Pennsylvania |
| | 209 S 33rd Street |
| | Philadelphia, PA 19104 |
| **Subject:** | Glycosyl linkage analysis. |
| **Sample** | Sample #1 |
| **CCRC Code:** | AR021617 |
| **Analyst:** | Ian Black |

**Methods:**

**Should any of these data be used in a publication, please include the following statement in the acknowledgment:  This work was supported by the Chemical Sciences, Geosciences and Biosciences Division, Office of Basic Energy Sciences, U.S. Department of Energy grant (DE-SC0015662)  to Parastoo Azadi " at the Complex Carbohydrate Research Center.**

<u>Glycosyl linkage analysis</u>

The glycosyl linkage analysis was performed at the Complex Carbohydrate Research Center and was supported by the Chemical Sciences, Geosciences and Biosciences Division, Office of Basic Energy Sciences, U.S. Department of Energy grant (DE-SC0015662)  to Parastoo Azadi.

# Glycosyl Linkage Analysis

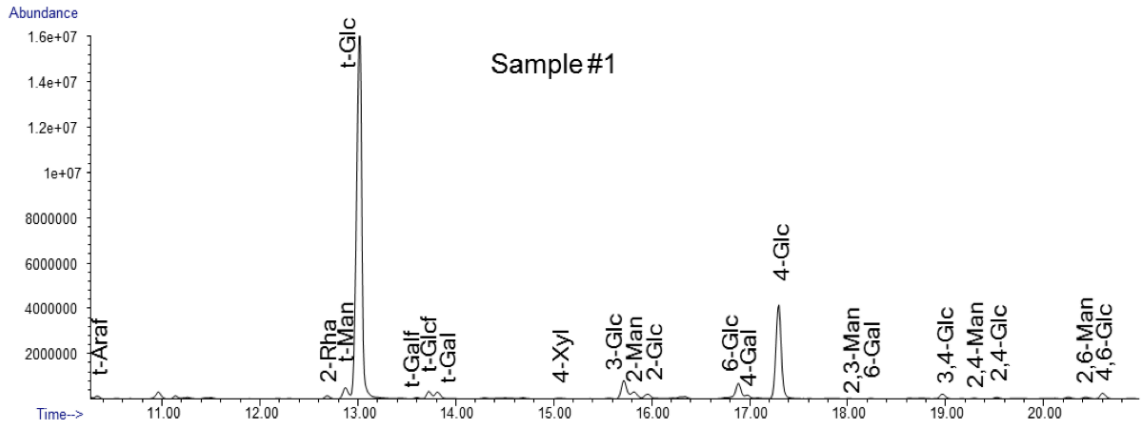

The GC chromatograms of the linkage analysis.

The relative percentage of each detected monosaccharide linkage in the sample.

| Peak | #1 area % |
|---|---|
| Terminal Arabinofuranosyl residue (t-Araf) | 0.4 |
| 2-linked Rhamnopyranosyl residue (2-Rha) | 0.5 |
| Terminal Mannopyranosyl residue (t-Man) | 2.1 |
| Terminal Glucopyranosyl residue (t-Glc) | 66.5 |
| Terminal Galactofuranosyl residue (t-Galf) | 0.2 |
| Terminal Glucofuranosyl residue (t-Glcf) | 1.2 |
| Terminal Galactopyranosyl residue (t-Gal) | 1.2 |
| 4-linked Xylopyranosyl residue (4-Xyl) | 0.1 |
| 3-linked Glucopyranosyl residue (3-Glc) | 3.2 |
| 2-linked Mannopyranosyl residue (2-Man) | 1.4 |
| 2-linked Glucopyranosyl residue (2-Glc) | 1.0 |
| 3-linked Galactopyranosyl residue (3-Gal) | - |
| 6-linked Glucopyranosyl residue (6-Glc) | 2.3 |
| 4-linked Galactopyranosyl residue (4-Gal) | 0.5 |
| 4-linked Glucopyranosyl residue (4-Glc) | 16.8 |
| 2,3-linked Mannopyranosyl residue (2,3-Man) | 0.1 |
| 6-linked Galactopyranosyl residue (6-Gal) | 0.1 |
| 3,4-linked Glucopyranosyl residue (3,4-Glc) | 0.8 |
| 2,4-linked Mannopyranosyl residue (2,4-Man) | 0.1 |
| 2,4-linked Glucopyranosyl residue (2,4-Glc) | 0.2 |
| 2,6-linked Mannopyranosyl residue (2,6-Man) | 0.3 |
| 4,6-linked Glucopyranosyl residue (4,6-Glc) | 1.0 |

**File S3: Nexus File**
```
#NEXUS
begin taxa;
	dimensions ntax=195;
	taxlabels
	10[&!name="gnetaceae",!color=#ff0000]
	101[&!name="cactaceae",!color=#0000ff]
	103[&!name="cornaceae",!color=#0000ff]
	104[&!name="loasaceae",!color=#0000ff]
	105[&!name="hydrangeaceae",!color=#0000ff]
	108[&!name="balsaminaceae",!color=#0000ff]
	11[&!name="welwitschiaceae",!color=#ff0000]
	111[&!name="lecythidaceae",!color=#0000ff]
	112[&!name="polemoniaceae",!color=#0000ff]
	113[&!name="fouquieriaceae",!color=#0000ff]
	114[&!name="sapotaceae",!color=#0000ff]
	115[&!name="ebenaceae",!color=#0000ff]
	116[&!name="primulaceae",!color=#0000ff]
	12[&!name="ephedraceae",!color=#ff0000]
	120[&!name="styracaceae",!color=#0000ff]
	122[&!name="sarraceniaceae",!color=#0000ff]
	123[&!name="actinidiaceae",!color=#0000ff]
	124[&!name="roridulaceae",!color=#0000ff]
	125[&!name="clethraceae",!color=#0000ff]
	127[&!name="ericaceae",!color=#0000ff]
	13[&!name="amborellaceae",!color=#400080]
	134[&!name="aquifoliaceae",!color=#0000ff]
	137[&!name="campanulaceae",!color=#0000ff]
	138[&!name="menyanthaceae",!color=#0000ff]
	140[&!name="asteraceae",!color=#0000ff]
	141[&!name="calyceraceae",!color=#0000ff]
	142[&!name="stylidiaceae",!color=#0000ff]
	15[&!name="cabombaceae",!color=#ff00ff]
	150[&!name="adoxaceae",!color=#0000ff]
	151[&!name="caprifoliaceae",!color=#0000ff]
	155[&!name="pittosporaceae",!color=#0000ff]
	156[&!name="araliaceae",!color=#0000ff]
	158[&!name="apiaceae",!color=#0000ff]
	16[&!name="nymphaeaceae",!color=#ff00ff]
	161[&!name="garryaceae",!color=#0000ff]
	165[&!name="rubiaceae",!color=#0000ff]
	166[&!name="gentianaceae",!color=#0000ff]
	168[&!name="gelsemiaceae",!color=#0000ff]
	169[&!name="apocynaceae",!color=#0000ff]
	17[&!name="austrobaileyaceae",!color=#cc66ff]
	170[&!name="boraginaceae",!color=#0000ff]
	171[&!name="convolvulaceae",!color=#0000ff]
	172[&!name="solanaceae",!color=#0000ff]
```

```
177[&!name="oleaceae",!color=#0000ff]
181[&!name="gesneriaceae",!color=#0000ff]
182[&!name="plantaginaceae",!color=#0000ff]
183[&!name="scrophulariaceae",!color=#0000ff]
184[&!name="stilbaceae",!color=#0000ff]
185[&!name="lamiaceae",!color=#0000ff]
186[&!name="phrymaceae",!color=#0000ff]
187[&!name="orobanchaceae",!color=#0000ff]
188[&!name="paulowniaceae",!color=#0000ff]
19[&!name="schisandraceae",!color=#cc66ff]
190[&!name="verbenaceae",!color=#0000ff]
191[&!name="bignoniaceae",!color=#0000ff]
193[&!name="martyniaceae",!color=#0000ff]
196[&!name="lentibulariaceae",!color=#0000ff]
198[&!name="acanthaceae",!color=#0000ff]
2[&!name="zamiaceae",!color=#ff0000]
200[&!name="paeoniaceae",!color=#0000ff]
202[&!name="hamamelidaceae",!color=#0000ff]
203[&!name="cercidiphyllaceae",!color=#0000ff]
206[&!name="grossulariaceae",!color=#0000ff]
207[&!name="saxifragaceae",!color=#0000ff]
208[&!name="crassulaceae",!color=#0000ff]
211[&!name="penthoraceae",!color=#0000ff]
212[&!name="haloragaceae",!color=#0000ff]
213[&!name="vitaceae",!color=#000080]
214[&!name="geraniaceae",!color=#000080]
215[&!name="melianthaceae",!color=#000080]
217[&!name="combretaceae",!color=#000080]
218[&!name="onagraceae",!color=#000080]
219[&!name="lythraceae",!color=#000080]
220[&!name="myrtaceae",!color=#000080]
222[&!name="melastomataceae",!color=#000080]
229[&!name="staphyleaceae",!color=#000080]
231[&!name="stachyuraceae",!color=#000080]
237[&!name="anacardiaceae",!color=#000080]
239[&!name="sapindaceae",!color=#000080]
24[&!name="aristolochiaceae",!color=#ff8000]
240[&!name="meliaceae",!color=#000080]
241[&!name="rutaceae",!color=#000080]
242[&!name="simaroubaceae",!color=#000080]
247[&!name="thymelaeaceae",!color=#000080]
248[&!name="malvaceae",!color=#000080]
249[&!name="cytinaceae",!color=#000080]
250[&!name="muntingiaceae",!color=#000080]
252[&!name="bixaceae",!color=#000080]
253[&!name="cistaceae",!color=#000080]
257[&!name="tropaeolaceae",!color=#000080]
26[&!name="piperaceae",!color=#ff8000]
```

```
261[&!name="limnanthaceae",!color=#000080]
267[&!name="resedaceae",!color=#000080]
268[&!name="tovariaceae",!color=#000080]
27[&!name="saururaceae",!color=#ff8000]
270[&!name="capparaceae",!color=#000080]
271[&!name="cleomaceae",!color=#000080]
272[&!name="brassicaceae",!color=#000080]
273[&!name="zygophyllaceae",!color=#000080]
276[&!name="fabaceae",!color=#000080]
278[&!name="polygalaceae",!color=#000080]
280[&!name="rosaceae",!color=#000080]
283[&!name="rhamnaceae",!color=#000080]
285[&!name="ulmaceae",!color=#000080]
286[&!name="cannabaceae",!color=#000080]
287[&!name="moraceae",!color=#000080]
288[&!name="urticaceae",!color=#000080]
289[&!name="coriariaceae",!color=#000080]
29[&!name="magnoliaceae",!color=#ff8000]
291[&!name="cucurbitaceae",!color=#000080]
293[&!name="begoniaceae",!color=#000080]
298[&!name="fagaceae",!color=#000080]
299[&!name="myricaceae",!color=#000080]
3[&!name="ginkgoaceae",!color=#ff0000]
300[&!name="juglandaceae",!color=#000080]
303[&!name="betulaceae",!color=#000080]
305[&!name="celastraceae",!color=#000080]
307[&!name="connaraceae",!color=#000080]
308[&!name="oxalidaceae",!color=#000080]
309[&!name="cunoniaceae",!color=#000080]
310[&!name="elaeocarpaceae",!color=#000080]
316[&!name="malpighiaceae",!color=#000080]
318[&!name="linaceae",!color=#000080]
32[&!name="annonaceae",!color=#ff8000]
321[&!name="calophyllaceae",!color=#000080]
322[&!name="hyperiacaceae",!color=#000080]
325[&!name="euphorbiaceae",!color=#000080]
327[&!name="phyllanthaceae",!color=#000080]
34[&!name="calycanthaceae",!color=#ff8000]
340[&!name="ochnaceae",!color=#000080]
342[&!name="achariaceae",!color=#000080]
343[&!name="violaceae",!color=#000080]
344[&!name="passifloraceae",!color=#000080]
347[&!name="salicaceae",!color=#000080]
352[&!name="dioscoreaceae",!color=#008040]
354[&!name="velloziaceae",!color=#008040]
357[&!name="cyclanthaceae",!color=#008040]
358[&!name="melanthiaceae",!color=#008040]
360[&!name="colchicaceae",!color=#008040]
```

```
361[&!name="alstroemeriaceae",!color=#008040]
364[&!name="smilacaceae",!color=#008040]
365[&!name="liliaceae",!color=#008040]
368[&!name="orchidaceae",!color=#008040]
375[&!name="tecophilaeaceae",!color=#008040]
376[&!name="doryanthaceae",!color=#008040]
377[&!name="iridaceae",!color=#008040]
379[&!name="xanthorrhoeaceae",!color=#008040]
380[&!name="amaryllidaceae",!color=#008040]
381[&!name="asparagaceae",!color=#008040]
382[&!name="arecaceae",!color=#008040]
384[&!name="commelinaceae",!color=#008040]
387[&!name="haemodoraceae",!color=#008040]
388[&!name="pontederiaceae",!color=#008040]
389[&!name="musaceae",!color=#008040]
390[&!name="heliconiaceae",!color=#008040]
392[&!name="lowiaceae",!color=#008040]
395[&!name="zingiberaceae",!color=#008040]
397[&!name="typhaceae",!color=#008040]
398[&!name="bromeliaceae",!color=#008040]
4[&!name="pinaceae",!color=#ff0000]
40[&!name="lauraceae",!color=#ff8000]
400[&!name="xyridaceae",!color=#008040]
410[&!name="poaceae",!color=#008040]
413[&!name="araceae",!color=#008040]
414[&!name="tofieldiaceae",!color=#008040]
415[&!name="alismataceae",!color=#008040]
416[&!name="hydrocharitaceae",!color=#008040]
417[&!name="butomaceae",!color=#008040]
419[&!name="aponogetonaceae",!color=#008040]
421[&!name="potamogetonaceae",!color=#008040]
425[&!name="juncaginaceae",!color=#008040]
43[&!name="papaveraceae",!color=#0080ff]
44[&!name="lardizabalaceae",!color=#0080ff]
47[&!name="berberidaceae",!color=#0080ff]
48[&!name="ranunculaceae",!color=#0080ff]
50[&!name="nelumbonaceae",!color=#66ffff]
51[&!name="platanaceae",!color=#66ffff]
53[&!name="trochodendraceae",!color=#0000ff]
54[&!name="buxaceae",!color=#0000ff]
56[&!name="gunneraceae",!color=#0000ff]
68[&!name="tamaricaceae",!color=#0000ff]
69[&!name="frankeniaceae",!color=#0000ff]
70[&!name="polygonaceae",!color=#0000ff]
71[&!name="plumbaginaceae",!color=#0000ff]
8[&!name="cupressaceae",!color=#ff0000]
81[&!name="amaranthaceae",!color=#0000ff]
83[&!name="caryophyllaceae",!color=#0000ff]
```

```
		88[&!name="aizoaceae",!color=#0000ff]
		9[&!name="taxaceae",!color=#ff0000]
		90[&!name="phytolaccaceae",!color=#0000ff]
		91[&!name="nyctaginaceae",!color=#0000ff]
		93[&!name="molluginaceae",!color=#0000ff]
		94[&!name="basellaceae",!color=#0000ff]
		98[&!name="talinaceae",!color=#0000ff]
		99[&!name="portulacaceae",!color=#0000ff]
	;
end;

begin trees;
	tree tree_1 = [&R]
(((3[&!color=#ff0000]:270.6651,((4[&!color=#ff0000]:171.1527,((8[
&!color=#ff0000]:64.3894,9[&!color=#ff0000]:64.3894)[&!color=#ff
0000]:106.7633)[&!color=#ff0000]:55.7774,((10[&!color=#ff0000]:9
1.7523,11[&!color=#ff0000]:91.7523)[&!color=#ff0000]:55.333,12[&
!color=#ff0000]:147.0853)[&!color=#ff0000]:79.8448)[&!color=#ff0
000]:43.735)[&!color=#ff0000]:22.1938,2[&!color=#ff0000]:292.858
9)[&!color=#ff0000]:33.7966,(13[&!color=#400080,!rotate=false]:1
83.9897,(((17[&!color=#cc66ff,!rotate=false]:95.8563,19[&!color=
#cc66ff,!rotate=false]:95.8563)[&!color=#cc66ff,!rotate=false]:7
3.6687,((((43[&!color=#0080ff,!rotate=true]:109.78,(44[&!color=#
0080ff,!rotate=true]:86.2052,(47[&!color=#0080ff,!rotate=true]:5
5.7161,48[&!color=#0080ff,!rotate=true]:55.7161)[&!color=#0080ff
,!rotate=true]:30.4891)[&!color=#0080ff,!rotate=true]:23.5748)[&
!color=#0080ff,!rotate=true]:35.6683,((50[&!color=#66ffff,!rotat
e=true]:102.0917,51[&!color=#66ffff,!rotate=true]:102.0917)[&!co
lor=#66ffff,!rotate=true]:37.8791,(53[&!color=#0000ff,!rotate=tr
ue]:136.2739,(((((213[&!color=#000080,!rotate=true]:121.8314,(((
(214[&!color=#000080,!rotate=true]:96.7591,215[&!color=#000080,!
rotate=true]:96.7591)[&!color=#000080,!rotate=true]:15.0635,(217
[&!color=#000080,!rotate=true]:90.6845,((218[&!color=#000080,!ro
tate=true]:57.3663,219[&!color=#000080,!rotate=true]:57.3663)[&!
color=#000080,!rotate=true]:31.8412,(220[&!color=#000080,!rotate
=true]:83.2571,222[&!color=#000080,!rotate=true]:83.2561)[&!colo
r=#000080,!rotate=true]:5.9513)[&!color=#000080,!rotate=true]:1.
477)[&!color=#000080,!rotate=true]:21.1371)[&!color=#000080,!rot
ate=true]:2.9317,((229[&!color=#000080,!rotate=true]:27.4086,231
[&!color=#000080,!rotate=true]:27.4086)[&!color=#000080,!rotate=
true]:84.2929,(((239[&!color=#000080,!rotate=true]:61.7208,(240[
&!color=#000080,!rotate=true]:51.8619,(241[&!color=#000080,!rota
te=true]:46.2672,242[&!color=#000080,!rotate=true]:46.2672)[&!co
lor=#000080,!rotate=true]:5.5947)[&!color=#000080,!rotate=true]:
9.8589)[&!color=#000080,!rotate=true]:1.6186,237[&!color=#000080
,!rotate=true]:63.3395)[&!color=#000080,!rotate=true]:38.3513,((
257[&!color=#000080,!rotate=true]:79.4339,(261[&!color=#000080,!
rotate=true]:67.784,((268[&!color=#000080,!rotate=true]:46.0205,
```

```
(270[&!color=#000080,!rotate=true]:30.3737,(271[&!color=#000080,!rotate=true]:22.1265,272[&!color=#000080,!rotate=true]:22.1265)[&!color=#000080,!rotate=true]:8.2481)[&!color=#000080,!rotate=true]:15.6469)[&!color=#000080,!rotate=true]:3.5921,267[&!color=#000080,!rotate=true]:49.6136)[&!color=#000080,!rotate=true]:18.1703)[&!color=#000080,!rotate=true]:11.6499)[&!color=#000080,!rotate=true]:11.0605,(247[&!color=#000080,!rotate=true]:67.3168,((248[&!color=#000080,!rotate=true]:56.5732,(249[&!color=#000080,!rotate=true]:39.743,250[&!color=#000080,!rotate=true]:39.743)[&!color=#000080,!rotate=true]:16.8302)[&!color=#000080,!rotate=true]:6.77,(252[&!color=#000080,!rotate=true]:55.2428,253[&!color=#000080,!rotate=true]:55.2428)[&!color=#000080,!rotate=true]:8.1004)[&!color=#000080,!rotate=true]:3.9737)[&!color=#000080,!rotate=true]:23.1765)[&!color=#000080,!rotate=true]:11.1974)[&!color=#000080,!rotate=true]:10.0108)[&!color=#000080,!rotate=true]:3.0517)[&!color=#000080,!rotate=true]:2.5592,((((((291[&!color=#000080,!rotate=true]:52.842,293[&!color=#000080,!rotate=true]:52.841)[&!color=#000080,!rotate=true]:2.9788,289[&!color=#000080,!rotate=true]:55.8208)[&!color=#000080,!rotate=true]:42.153,(298[&!color=#000080,!rotate=true]:58.0181,((299[&!color=#000080,!rotate=true]:28.539,300[&!color=#000080,!rotate=true]:28.539)[&!color=#000080,!rotate=true]:13.4866,303[&!color=#000080,!rotate=true]:42.0256)[&!color=#000080,!rotate=true]:15.9935)[&!color=#000080,!rotate=true]:39.9556)[&!color=#000080,!rotate=true]:3.5135,(280[&!color=#000080,!rotate=true]:83.2662,((285[&!color=#000080,!rotate=true]:56.5445,(286[&!color=#000080,!rotate=true]:44.2088,(287[&!color=#000080,!rotate=true]:34.5194,288[&!color=#000080,!rotate=true]:34.5194)[&!color=#000080,!rotate=true]:9.6894)[&!color=#000080,!rotate=true]:12.3357)[&!color=#000080,!rotate=true]:11.9465,283[&!color=#000080,!rotate=true]:68.49)[&!color=#000080,!rotate=true]:14.7762)[&!color=#000080,!rotate=true]:18.22)[&!color=#000080,!rotate=true]:1.9956,(276[&!color=#000080,!rotate=true]:59.3632,278[&!color=#000080,!rotate=true]:59.3632)[&!color=#000080,!rotate=true]:44.1196)[&!color=#000080,!rotate=true]:8.1179,(((((327[&!color=#000080,!rotate=true]:86.2324,(340[&!color=#000080,!rotate=true]:72.3138,(342[&!color=#000080,!rotate=true]:59.3003,((343[&!color=#000080,!rotate=true]:46.7128,344[&!color=#000080,!rotate=true]:46.7128)[&!color=#000080,!rotate=true]:10.9498,347[&!color=#000080,!rotate=true]:57.6626)[&!color=#000080,!rotate=true]:1.6377)[&!color=#000080,!rotate=true]:13.0135)[&!color=#000080,!rotate=true]:13.9185)[&label="B",!color=#000080,!rotate=true]:11.4435,325[&!color=#000080,!rotate=true]:97.6758)[&!color=#000080,!rotate=true]:2.8372,((318[&!color=#000080,!rotate=true]:94.9126,(321[&!color=#000080,!rotate=true]:87.0564,322[&!color=#000080,!rotate=true]:87.0564)[&!color=#000080,!rotate=true]:7.8562)[&!color=#000080,!rotate=true]:2.96,316[&!color=#000080,!rotate=true]:97.8726)[&!color=#000080,!rotate=true]:2.6414)[&!color=#000080,!rotate=true]:5.8939,((307[&!color=#000080,!rota
```

te=true]:36.8778,308[&!color=#000080,!rotate=true]:36.8778)[&!color=#000080,!rotate=true]:24.7964,(309[&!color=#000080,!rotate=true]:47.4203,310[&!color=#000080,!rotate=true]:47.4203)[&!color=#000080,!rotate=true]:14.2539)[&!color=#000080,!rotate=true]:44.7346)[&!color=#000080,!rotate=true]:1.077,305[&!color=#000080,!rotate=true]:107.4848)[&!color=#000080,!rotate=true]:4.1159)[&!color=#000080,!rotate=true]:1.858,273[&!color=#000080,!rotate=true]:113.4587)[&!color=#000080,!rotate=true]:3.8537)[&!color=#000080,!rotate=true]:4.52)[&!color=#000080,!rotate=true]:1.7341,((200[&!color=#0000ff,!rotate=true]:80.5335,(202[&!color=#0000ff,!rotate=true]:60.0028,203[&!color=#0000ff,!rotate=true]:60.0018)[&!color=#0000ff,!rotate=true]:20.5307)[&!color=#0000ff,!rotate=true]:3.2827,((208[&!color=#0000ff,!rotate=true]:64.6079,(211[&!color=#0000ff,!rotate=true]:30.4605,212[&!color=#0000ff,!rotate=true]:30.4605)[&!color=#0000ff,!rotate=true]:34.1464)[&!color=#0000ff,!rotate=true]:15.2555,(206[&!color=#0000ff,!rotate=true]:43.8922,207[&!color=#0000ff,!rotate=true]:43.8922)[&!color=#0000ff,!rotate=true]:35.9702)[&!color=#0000ff,!rotate=true]:3.9528)[&!color=#0000ff,!rotate=true]:39.7493)[&!color=#0000ff,!rotate=true]:2.2604,((((68[&!color=#0000ff,!rotate=true]:43.4508,69[&!color=#0000ff,!rotate=true]:43.4508)[&!color=#0000ff,!rotate=true]:36.4343,(70[&!color=#0000ff,!rotate=true]:48.7324,71[&!color=#0000ff,!rotate=true]:48.7324)[&!color=#0000ff,!rotate=true]:31.1527)[&!color=#0000ff,!rotate=true]:14.6146,((81[&!color=#0000ff,!rotate=true]:53.1362,83[&!color=#0000ff,!rotate=true]:53.1372)[&!color=#0000ff,!rotate=true]:13.344,((93[&!color=#0000ff,!rotate=true]:47.1586,(94[&!color=#0000ff,!rotate=true]:36.015,(98[&!color=#0000ff,!rotate=true]:25.1098,(99[&!color=#0000ff,!rotate=true]:21.1636,101[&!color=#0000ff,!rotate=true]:21.1636)[&!color=#0000ff,!rotate=true]:3.9462)[&!color=#0000ff,!rotate=true]:10.9053)[&!color=#0000ff,!rotate=true]:11.1436)[&!color=#0000ff,!rotate=true]:7.765,(88[&!color=#0000ff,!rotate=true]:35.282,(90[&!color=#0000ff,!rotate=true]:28.2974,91[&!color=#0000ff,!rotate=true]:28.2974)[&!color=#0000ff,!rotate=true]:6.9847)[&!color=#0000ff,!rotate=true]:19.6415)[&!color=#0000ff,!rotate=true]:11.5567)[&!color=#0000ff,!rotate=true]:28.0204)[&!color=#0000ff,!rotate=true]:19.9598,((((111[&!color=#0000ff,!rotate=true]:69.4548,((112[&!color=#0000ff,!rotate=true]:54.0609,113[&!color=#0000ff,!rotate=true]:54.0609)[&!color=#0000ff,!rotate=true]:12.2736,((114[&!color=#0000ff,!rotate=true]:59.141,(115[&!color=#0000ff,!rotate=true]:53.0118,116[&!color=#0000ff,!rotate=true]:53.0118)[&!color=#0000ff,!rotate=true]:6.1292)[&!color=#0000ff,!rotate=true]:3.8985,(120[&!color=#0000ff,!rotate=true]:56.6167,((122[&!color=#0000ff,!rotate=true]:39.3478,(123[&!color=#0000ff,!rotate=true]:30.8951,124[&!color=#0000ff,!rotate=true]:30.8951)[&!color=#0000ff,!rotate=true]:8.4527)[&!color=#0000ff,!rotate=true]:7.2371,(125[&!color=#0000ff,!rotate=true]:38.7437,127[&!color=#0000ff,!rotate=true]:38.7437)[&!color=#0000ff,!rotate=true]:7.8412)[&!color=#0000

```
ff,!rotate=true]:10.0318)[&!color=#0000ff,!rotate=true]:6.4227)[&!color=#0000ff,!rotate=true]:3.2941)[&!color=#0000ff,!rotate=true]:3.1213)[&!color=#0000ff,!rotate=true]:23.515,108[&!color=#0000ff,!rotate=true]:92.9698)[&!color=#0000ff,!rotate=true]:11.2891,(((((138[&!color=#0000ff,!rotate=true]:40.3522,(140[&!color=#0000ff,!rotate=true]:18.7465,141[&!color=#0000ff,!rotate=true]:18.7465)[&!color=#0000ff,!rotate=true]:21.6056)[&!color=#0000ff,!rotate=true]:8.9609,142[&!color=#0000ff,!rotate=true]:49.3131)[&!color=#0000ff,!rotate=true]:9.457,137[&!color=#0000ff,!rotate=true]:58.7701)[&!color=#0000ff,!rotate=true]:17.2638,((150[&!color=#0000ff,!rotate=true]:41.4061,151[&!color=#0000ff,!rotate=true]:41.4061)[&!color=#0000ff,!rotate=true]:21.0377,(155[&!color=#0000ff,!rotate=true]:29.4913,(156[&!color=#0000ff,!rotate=true]:26.7144,158[&!color=#0000ff,!rotate=true]:26.7144)[&!color=#0000ff,!rotate=true]:2.7768)[&!color=#0000ff,!rotate=true]:32.9535)[&!color=#0000ff,!rotate=true]:13.5892)[&!color=#0000ff,!rotate=true]:16.4615,134[&!color=#0000ff,!rotate=true]:92.4954)[&!color=#0000ff,!rotate=true]:5.0256,(161[&!color=#0000ff,!rotate=true]:89.5513,((165[&!color=#0000ff,!rotate=true]:56.0489,(166[&!color=#0000ff,!rotate=true]:43.5349,(168[&!color=#0000ff,!rotate=true]:34.0727,169[&!color=#0000ff,!rotate=true]:34.0727)[&!color=#0000ff,!rotate=true]:9.4622)[&!color=#0000ff,!rotate=true]:12.514)[&!color=#0000ff,!rotate=true]:15.5422,(170[&!color=#0000ff,!rotate=true]:68.0705,((171[&!color=#0000ff,!rotate=true]:37.4597,172[&!color=#0000ff,!rotate=true]:37.4597)[&!color=#0000ff,!rotate=true]:29.0688,(177[&!color=#0000ff,!rotate=true]:50.9547,((182[&!color=#0000ff,!rotate=true]:38.6666,(183[&!color=#0000ff,!rotate=true]:36.0308,(184[&!color=#0000ff,!rotate=true]:34.4251,((185[&!color=#0000ff,!rotate=true]:29.4367,(186[&!color=#0000ff,!rotate=true]:26.8774,(187[&!color=#0000ff,!rotate=true]:22.112,188[&!color=#0000ff,!rotate=true]:22.112)[&!color=#0000ff,!rotate=true]:4.7654)[&!color=#0000ff,!rotate=true]:2.5593)[&!color=#0000ff,!rotate=true]:3.1899,((((196[&!color=#0000ff,!rotate=true]:25.9174,198[&!color=#0000ff,!rotate=true]:25.9174)[&!color=#0000ff,!rotate=true]:2.5591,193[&!color=#0000ff,!rotate=true]:28.4765)[&!color=#0000ff,!rotate=true]:1.1342,191[&!color=#0000ff,!rotate=true]:29.6107)[&!color=#0000ff,!rotate=true]:1.079,190[&!color=#0000ff,!rotate=true]:30.6896)[&!color=#0000ff,!rotate=true]:1.937)[&!color=#0000ff,!rotate=true]:1.7985)[&!color=#0000ff,!rotate=true]:1.6057)[&!color=#0000ff,!rotate=true]:2.6358)[&!color=#0000ff,!rotate=true]:1.3669,181[&!color=#0000ff,!rotate=true]:40.0325)[&!color=#0000ff,!rotate=true]:10.9212)[&!color=#0000ff,!rotate=true]:15.5748)[&!color=#0000ff,!rotate=true]:1.541)[&!color=#0000ff,!rotate=true]:3.5216)[&!color=#0000ff,!rotate=true]:17.9592)[&!color=#0000ff,!rotate=true]:7.9707)[&!color=#0000ff,!rotate=true]:6.7379)[&!color=#0000ff,!rotate=true]:1.7852,(103[&!color=#0000ff,!rotate=true]:69.2055,(104[&!color=#0000ff,!rotate=true]:38.3012,105[&!color=#0000ff,!rotate=true]:38.3012)[&!color=
```

```
#0000ff,!rotate=true]:30.9033)[&!color=#0000ff,!rotate=true]:36.8386)[&!color=#0000ff,!rotate=true]:8.4153)[&!color=#0000ff,!rotate=true]:11.3665)[&!color=#0000ff,!rotate=true]:2.7873,56[&!color=#0000ff,!rotate=true]:128.6132)[&!color=#0000ff,!rotate=true]:5.5109,54[&!color=#0000ff,!rotate=true]:134.1241)[&!color=#0000ff,!rotate=true]:2.1498)[&!color=#0000ff,!rotate=true]:3.6969)[&!color=#0000ff,!rotate=true]:5.4775)[&!color=#0000ff,!rotate=true]:14.2652,((413[&!color=#008040,!rotate=true]:121.5893,(414[&!color=#008040,!rotate=true]:116.0839,((415[&!color=#008040,!rotate=true]:72.877,(416[&!color=#008040,!rotate=true]:65.5031,417[&!color=#008040,!rotate=true]:65.5031)[&!color=#008040,!rotate=true]:7.3739)[&!color=#008040,!rotate=true]:21.7587,(419[&!color=#008040,!rotate=true]:81.4526,(425[&!color=#008040,!rotate=true]:63.1043,421[&!color=#008040,!rotate=true]:63.1043)[&!color=#008040,!rotate=true]:18.3483)[&!color=#008040,!rotate=true]:13.182)[&!color=#008040,!rotate=true]:21.4483)[&!color=#008040,!rotate=true]:5.5054)[&!color=#008040,!rotate=true]:13.9164,((352[&!color=#008040,!rotate=true]:92.9267,(354[&!color=#008040,!rotate=true]:57.9442,357[&!color=#008040,!rotate=true]:57.9452)[&!color=#008040,!rotate=true]:34.9825)[&!color=#008040,!rotate=true]:25.2033,(((368[&!color=#008040,!rotate=true]:99.0365,((376[&!color=#008040,!rotate=true]:62.2398,(377[&!color=#008040,!rotate=true]:52.5609,(379[&!color=#008040,!rotate=true]:34.6323,(380[&!color=#008040,!rotate=true]:28.5824,381[&!color=#008040,!rotate=true]:28.5824)[&!color=#008040,!rotate=true]:6.0499)[&!color=#008040,!rotate=true]:17.9287)[&!color=#008040,!rotate=true]:9.6788)[&!color=#008040,!rotate=true]:3.0008,375[&!color=#008040,!rotate=true]:65.2416)[&!color=#008040,!rotate=true]:33.7949)[&!color=#008040,!rotate=true]:12.9637,((382[&!color=#008040,!rotate=true]:104.2124,((384[&!color=#008040,!rotate=true]:76.1336,(387[&!color=#008040,!rotate=true]:55.6399,388[&!color=#008040,!rotate=true]:55.6399)[&!color=#008040,!rotate=true]:20.4938)[&!color=#008040,!rotate=true]:17.6002,(389[&!color=#008040,!rotate=true]:84.6442,((390[&!color=#008040,!rotate=true]:61.4558,392[&!color=#008040,!rotate=true]:61.4558)[&!color=#008040,!rotate=true]:12.7915,395[&!color=#008040,!rotate=true]:74.2473)[&!color=#008040,!rotate=true]:10.3969)[&!color=#008040,!rotate=true]:9.0887)[&!color=#008040,!rotate=true]:10.4796)[&!color=#008040,!rotate=true]:2.453,((397[&!color=#008040,!rotate=true]:79.4529,398[&!color=#008040,!rotate=true]:79.4529)[&!color=#008040,!rotate=true]:10.829,(400[&!color=#008040,!rotate=true]:81.8867,410[&!color=#008040,!rotate=true]:81.8877)[&!color=#008040,!rotate=true]:8.3951)[&!color=#008040,!rotate=true]:16.3826)[&!color=#008040,!rotate=true]:5.3348)[&!color=#008040,!rotate=true]:3.5025,((358[&!color=#008040,!rotate=true]:81.4059,(360[&!color=#008040,!rotate=true]:51.561,361[&!color=#008040,!rotate=true]:51.561)[&!color=#008040,!rotate=true]:29.8449)[&!color=#008040,!rotate=true]:5.6129,(364[&!color=#008040,!rotate=true]:40.9678,365[&!color=#008040,!rotate=true]:
```

```
40.9678)[&!color=#008040,!rotate=true]:46.05)[&!color=#008040,!r
otate=true]:28.4839)[&!color=#008040,!rotate=true]:2.6273)[&!col
or=#008040,!rotate=true]:17.3756)[&!color=#008040,!rotate=true]:
24.2089)[&!rotate=true]:1.059,(((34[&!color=#ff8000,!rotate=true
]:110.6748,40[&!color=#ff8000,!rotate=true]:110.6748)[&!color=#f
f8000,!rotate=true]:13.5402,(29[&!color=#ff8000,!rotate=true]:58
.2596,32[&!color=#ff8000,!rotate=true]:58.2596)[&!color=#ff8000,
!rotate=true]:65.9554)[&!color=#ff8000,!rotate=true]:9.4618,((26
[&!color=#ff8000,!rotate=true]:54.2685,27[&!color=#ff8000,!rotat
e=true]:54.2685)[&!color=#ff8000,!rotate=true]:48.4033,24[&!colo
r=#ff8000,!rotate=true]:102.6718)[&!color=#ff8000,!rotate=true]:
31.005)[&!color=#ff8000,!rotate=true]:27.0967)[&!rotate=true]:8.
7515)[&!rotate=false]:8.6069,(15[&!color=#ff00ff,!rotate=true]:3
2.54,16[&!color=#ff00ff,!rotate=true]:32.54)[&!color=#ff00ff,!ro
tate=true]:145.5919)[&!rotate=true]:5.8577)[&!rotate=false]:142.
6659);
end;

begin figtree;
	set appearance.backgroundColorAttribute="Default";
	set appearance.backgroundColour=#ffffff;
	set appearance.branchColorAttribute="User selection";
	set appearance.branchColorGradient=false;
	set appearance.branchLineWidth=1.0;
	set appearance.branchMinLineWidth=0.0;
	set appearance.branchWidthAttribute="Fixed";
	set appearance.foregroundColour=#000000;
	set appearance.hilightingGradient=false;
	set appearance.selectionColour=#2d3680;
	set branchLabels.colorAttribute="User selection";
	set branchLabels.displayAttribute="Branch times";
	set branchLabels.fontName="sansserif";
	set branchLabels.fontSize=8;
	set branchLabels.fontStyle=0;
	set branchLabels.isShown=false;
	set branchLabels.significantDigits=4;
	set layout.expansion=0;
	set layout.layoutType="RECTILINEAR";
	set layout.zoom=0;
	set legend.attribute="label";
	set legend.fontSize=10.0;
	set legend.isShown=false;
	set legend.significantDigits=4;
	set nodeBars.barWidth=4.0;
	set nodeBars.displayAttribute=null;
	set nodeBars.isShown=false;
	set nodeLabels.colorAttribute="User selection";
	set nodeLabels.displayAttribute="Node ages";
```

```
set nodeLabels.fontName="sansserif";
set nodeLabels.fontSize=8;
set nodeLabels.fontStyle=0;
set nodeLabels.isShown=false;
set nodeLabels.significantDigits=4;
set nodeShapeExternal.colourAttribute="User selection";
set nodeShapeExternal.isShown=false;
set nodeShapeExternal.minSize=10.0;
set nodeShapeExternal.scaleType=Width;
set nodeShapeExternal.shapeType=Circle;
set nodeShapeExternal.size=4.0;
set nodeShapeExternal.sizeAttribute="Fixed";
set nodeShapeInternal.colourAttribute="User selection";
set nodeShapeInternal.isShown=false;
set nodeShapeInternal.minSize=10.0;
set nodeShapeInternal.scaleType=Width;
set nodeShapeInternal.shapeType=Circle;
set nodeShapeInternal.size=4.0;
set nodeShapeInternal.sizeAttribute="Fixed";
set polarLayout.alignTipLabels=false;
set polarLayout.angularRange=0;
set polarLayout.rootAngle=0;
set polarLayout.rootLength=100;
set polarLayout.showRoot=true;
set radialLayout.spread=0.0;
set rectilinearLayout.alignTipLabels=false;
set rectilinearLayout.curvature=0;
set rectilinearLayout.rootLength=100;
set scale.offsetAge=0.0;
set scale.rootAge=1.0;
set scale.scaleFactor=1.0;
set scale.scaleRoot=false;
set scaleAxis.automaticScale=true;
set scaleAxis.fontSize=8.0;
set scaleAxis.isShown=false;
set scaleAxis.lineWidth=1.0;
set scaleAxis.majorTicks=1.0;
set scaleAxis.minorTicks=0.5;
set scaleAxis.origin=0.0;
set scaleAxis.reverseAxis=false;
set scaleAxis.showGrid=true;
set scaleBar.automaticScale=true;
set scaleBar.fontSize=10.0;
set scaleBar.isShown=true;
set scaleBar.lineWidth=1.0;
set scaleBar.scaleRange=0.0;
set tipLabels.colorAttribute="User selection";
set tipLabels.displayAttribute="Names";
```

```
        set tipLabels.fontName="sansserif";
        set tipLabels.fontSize=7;
        set tipLabels.fontStyle=0;
        set tipLabels.isShown=true;
        set tipLabels.significantDigits=4;
        set trees.order=false;
        set trees.orderType="increasing";
        set trees.rooting=false;
        set trees.rootingType="User Selection";
        set trees.transform=false;
        set trees.transformType="cladogram";
end;
```